\documentclass[aps,11pt,showkeys,nofootinbib]{revtex4-1}

\usepackage{amsmath,amssymb,amsfonts,amsthm}
\usepackage{amsbsy} 
\usepackage{epsfig}
\usepackage{latexsym}

\usepackage{hyperref}

\newcommand{\oarX}[1]{\href{http://arxiv.org/abs/#1}{{\ttfamily #1}}}
\newcommand{\arX}[1]{\href{http://arxiv.org/abs/#1}{{\ttfamily arXiv:#1}}}
\newcommand{\doin}[2]{\href{http://dx.doi.org/#1}{#2}}

\def\be{\begin{equation}}
\def\ee{\end{equation}}
\def\dd{{\rm d}}
\def\bes{\begin{eqnarray}}
\def\ees{\end{eqnarray}}

\DeclareMathOperator{\tr}{tr}

\newcommand{\R}{\mathbb{R}}
\newcommand{\C}{\mathbb{C}}
\newcommand{\ii}{\text{i}}
\newcommand{\dee}{\text{d}}
\newcommand{\pd}[2]{ \frac{ \partial #1 }{ \partial #2 } }
\newcommand{\mat}[1]{ \begin{pmatrix} #1 \end{pmatrix} }

\renewcommand{\bar}{\overline}

\begin{document}

\title{Quantum cosmology of pure connection general relativity}
\author{Steffen Gielen}
\email{s.c.gielen@sheffield.ac.uk}
\author{Elliot Nash}
\email{enash4@sheffield.ac.uk}
\affiliation{School of Mathematics and Statistics, University of Sheffield, Hicks Building, Hounsfield Road, Sheffield S3 7RH, United Kingdom}
\date{\today}

\begin{abstract}
We study homogeneous cosmological models in formulations of general relativity with cosmological constant based on a (complexified) connection rather than a spacetime metric, in particular in a first order theory obtained by integrating out the self-dual two-forms in the chiral Pleba\'{n}ski formulation. Classical dynamics for the Bianchi IX model are studied in the Lagrangian and Hamiltonian formalism, where we emphasise the reality conditions needed to obtain real Lorentzian solutions. The solutions to these reality conditions fall into different branches, which in turn lead to different real Hamiltonian theories, only one of which is the usual Lorentzian Bianchi IX model. We also show the simpler case of the flat Bianchi I model, for which both the reality conditions and dynamical equations simplify considerably. We discuss the relation of a real Euclidean version of the same theory to this complex theory. Finally, we study the quantum theory of homogeneous and isotropic models, for which the pure connection action for general relativity reduces to a pure boundary term and the path integral is evaluated immediately, reproducing known results in quantum cosmology. An intriguing aspect of these theories is that the signature of the effective spacetime metric, and hence the interpretation of the cosmological constant, are intrinsically ambiguous.
\end{abstract}

\maketitle

\tableofcontents

\section{Introduction}

General relativity is often thought of as a theory of Lorentzian metrics satisfying Einstein's equations, with the affine connection fixed to be the metric-compatible, torsion-free (Levi-Civita) connection for a given metric. However, there are many different, classically more or less equivalent ways of defining the same theory. One can go from a diffeomorphism-invariant theory to one with larger or smaller symmetry group \cite{MoreorLess}; one can introduce a local Lorentz gauge symmetry and work with a spin connection and orthonormal frame field as independent variables (possibly adding torsion); one can work with complex self-dual variables encoding the metric and spin connection~\cite{Plebanski,PlebanskiIntro}; or one can encode the effects of gravity into torsion or non-metricity. Even when these formulations agree classically, they may suggest different avenues towards quantisation, which is one of the main motivations for exploring different formulations. As an example, promising results towards canonical quantisation were achieved in loop quantum gravity~\cite{LQGreviews} once a reformulation of general relativity in terms of Ashtekar variables had been constructed~\cite{Ashtekar}. Many different formulations of general relativity and some generalised gravitational theories are reviewed in~\cite{peldan}, mostly following the general philosophy of trying to replace the metric  as the fundamental field by a connection. This could help to bring gravity closer to the language of the other forces in Nature, and suggest new avenues for unification of gravity and other forces~\cite{KrasnovPercacci}. 

Building on earlier attempts such as~\cite{CDJ}, one possible endpoint of this strategy to remove the metric from the theory was reached when a ``pure connection'' formulation of general relativity was found in~\cite{KrasnovPureConn}. The pure connection formulation can be obtained from the Pleba\'{n}ski formulation~\cite{Plebanski} of general relativity after the two-form fields encoding the metric have been ``integrated out,'' leaving in the minimal formulation an action that only depends on the (complex) spin connection. Intermediate forms exist in which the action no longer depends on a metric or frame field, but still contains some auxiliary fields. A comprehensive overview over these theories, their historical development and mathematical structure is given in~\cite{KrasnovBook}. The main goal of this paper is to explore the dynamical structure of these theories in the simplest nontrivial setting of homogeneous cosmology, both in Lagrangian and Hamiltonian form, and to take some steps towards quantisation.

In the continued absence of a fully satisfactory ``quantisation of gravity,'' cosmological models have long been employed as a starting point for studying technical and conceptual issues in quantum gravity, be it in the original metric formulation~\cite{misner}, loop quantum gravity~\cite{LQC}, or discrete approaches such as Regge calculus~\cite{ReggeQC}. They might also be seen as a starting point for uncovering possible observational signatures of a quantisation of gravity~\cite{HalliwellHawking}. While the study of quantum cosmology is an old subject, basic questions such as the definition of the path integral for such models are still subject to debate~\cite{QCdebate}; for instance, one may ask which {\em complex} metrics should be summed over in a semiclassical approximation~\cite{complexmetrics}.  We will revisit the quantum cosmology path integral from the perspective of a connection-based formulation. The question of real or complex configurations takes a different form in the Pleba\'{n}ski formulation and its extensions: here, even classically the ``Lorentzian'' version of the theory is naturally complex, and reality conditions need be imposed to restrict to solutions interpreted as Lorentzian. We will show that in spatially homogeneous (anisotropic) models these reality conditions can already be quite nontrivial and admit various solution branches, only some of which correspond to the Lorentzian solutions of interest. An alternative is to start with a Euclidean version of the theory which is manifestly real, and obtain Lorentzian solutions from such a theory by ``analytic continuation,'' with all the ambiguities such a procedure would entail once one generalises beyond homogeneity. In either Euclidean or Lorentzian versions, an interesting subtlety of the connection-based theories we study is that the (Urbantke) metrics associated to different field configurations can have different signatures. We will see that this leads to an ambiguity in interpreting the sign of the cosmological constant.

Many of the significant technical difficulties encountered in attempts to quantise gravity are related to the symmetry of gravity under spacetime diffeomorphisms; the structure of these symmetry transformations is significantly more complicated than that of, e.g., gauge transformations in Yang--Mills theory. In the path integral setting, this symmetry leads to a huge redundancy of the formally defined integral over all metrics\footnote{An alternative is to use a discrete definition of the path integral, such as in causal dynamical triangulations~\cite{CDTreview}.}. At the level of spatially homogeneous cosmology, the remnant of diffeomorphism symmetry is a symmetry under reparametrisations of the time coordinate, which can be implemented in the path integral using the BFV (Batalin--Fradkin--Vilkovisky) formalism~\cite{Halliwell}. After gauge fixing one is generally left with an ordinary integral over the proper time between initial and final state, which can be approximated by a sum over complex saddle points (see, e.g.,~\cite{LorentzianQC}). An important role is played by the lapse function, which has to transform nontrivially under reparametrisations of time in order to ``compensate'' for the transformation of other fields. The same symmetry appears differently in pure connection formulations, which have no lapse function and are invariant under reparametrisations without such compensating transformations. We will see that the difference between these formulations is analogous to the relation between different actions for a relativistic particle which may or may not contain additional Lagrange multiplier fields, and hence suggest different Lagrangian definitions for the path integral in the quantum theory, but lead to equivalent Hamiltonian formulations. 

One particularly interesting observation in our analysis is that the action for the simplest pure connection formulation of~\cite{KrasnovPureConn}, when restricted to homogeneous and isotropic FLRW (Friedmann--Lema\^{i}tre--Robertson--Walker) geometries, reduces to a pure boundary term. Hence, the naive (Lagrangian) path integral is evaluated immediately, given that the value of the action does not depend on the histories summed over; one only needs to deal with the redundancy of integrating over equivalent configurations. The result of such a pure connection path integral is the analogue of the Chern--Simons state restricted to FLRW geometries, which is in some sense dual to the Hartle--Hawking/Vilenkin wavefunctions in the metric formulation~\cite{MagueijoDuality}. From the phase space perspective, it corresponds to a path integral in which the connection, but not the metric, is specified at the initial and final times (see also \cite{RayJoao} for a very recent discussion of this). Apart from conceptual insights, this observation shows how to reproduce known results in quantum cosmology in simpler terms from a connection representation of general relativity, which may prove useful in the quantisation of models with less symmetry.

The structure of the paper is as follows. After discussing the simpler case of different actions for a  relativistic particle, section \ref{actionsec} reviews different actions for general relativity in which the main variable is an $SO(3,\C)$ connection rather than the metric. Starting with the Pleba\'{n}ski formulation in terms of a self-dual connection and self-dual two-forms, we review how to obtain an intermediate {\em first order chiral connection action} in which the two-forms have disappeared, and then a pure connection formulation. All these theories are intrinsically complex, and suitable reality conditions need to be imposed. Spatially homogeneous (Bianchi) models are introduced in section \ref{Homogeneity section}. Our new results start in section \ref{Diagonal symmetric models section}, where we discuss the Bianchi IX model in the chiral first order theory, extending the analysis of \cite{KrasnovBook}. We show that a symmetry-reduced action gives the dynamics for spatially homogeneous fields in the full theory, and propose a complex (holomorphic) Hamiltonian formulation. While the theory does not contain a fundamental spacetime metric, we use the Urbantke metric to characterise real Lorentzian solutions. There are distinct branches of Lorentzian solutions, out of which one is the standard Lorentzian Bianchi IX model and the others correspond to a metric with timelike surfaces of homogeneity. We give a detailed construction of a Hamiltonian system for the most relevant Lorentzian branch, showing that this is equivalent to the usual Lorentzian Bianchi IX model. In section \ref{Bianchi I section} we take the limit of vanishing spatial curvature and obtain the dynamics of the Bianchi I model, where analytical solutions can be found more easily. In section \ref{euclideansec} we discuss a Euclidean version of the chiral connection theory, which only uses real $SO(3)$ variables and does not require reality conditions. We show how to map solutions to this theory to all of the Lorentzian branches using complex transformations which may be interpreted as a Wick rotation. Finally, in section \ref{homoiso} we restrict to FLRW symmetry. For this case, the path integral with connection boundary conditions can be evaluated immediately. We show that this corresponds to the fact that the classical pure connection action for this model is a boundary term, so that there are no dynamical equations. We close with some concluding remarks and an outlook.

\section{Actions for general relativity}
\label{actionsec}

\subsection{Partial analogy: relativistic particle actions}

We will study different actions for vacuum general relativity (with cosmological constant) related to each other by classically ``integrating out'' some dynamical variables, i.e., substituting solutions to the equations of motion back into the action. The actions are all invariant with respect to arbitrary coordinate changes $x^\mu\mapsto\tilde{x}^\nu=\tilde{x}^\nu(x^\mu)$ or, in the case of homogeneous cosmological models, under reparametrisation of the time coordinate, $t\mapsto \tilde{t}=\tilde{t}(t)$, but the way in which this transformation leaves the action invariant differs between different formulations. These points can be illustrated using a simpler example, namely actions for a relativistic particle in flat spacetime.

The most common way of defining such an action is to demand that the proper length of the particle worldline be extremal; that is, one starts from the action
\be
S[x^\mu,\dot{x}^\mu]= - m\int \dd\tau \sqrt{-\eta_{\mu\nu}\dot{x}^\mu\dot{x}^\nu}
\label{particleaction1}
\ee 
where $x^\mu(\tau)$ are the Minkowski spacetime coordinates of the particle parametrised by an arbitrary worldline parameter $\tau$, $\dot{}$ denotes derivative with respect to $\tau$, and $m$ is the particle mass. (\ref{particleaction1}) is invariant under a reparametrisation
\be
\tau\mapsto\tilde{\tau}=\tilde{\tau}(\tau)\,,\quad \dd\tau\mapsto \dd\tilde\tau=\dd\tau\,\tilde{\tau}'(\tau)\,,\quad \frac{\dd x^\mu}{\dd \tau}\mapsto  \frac{\dd x^\mu}{\dd \tilde\tau}=\frac{\dd x^\mu}{\dd \tau}(\tilde{\tau}'(\tau))^{-1}
\ee
where we assume that $\tilde{\tau}'(\tau)>0$ everywhere so that the reparametrisation is well-defined everywhere and preserves the orientation of time, and we write out the $\tau$ and $\tilde{\tau}$ derivatives explicitly for clarity. At the Hamiltonian level, this gauge symmetry leads to a primary constraint, since we have
\be
p_\mu := \frac{\partial\mathcal{L}}{\partial \dot{x}^\mu} =  \frac{m\eta_{\mu\nu}\dot{x}^\nu}{\sqrt{-\eta_{\mu\nu}\dot{x}^\mu\dot{x}^\nu}}\qquad \Rightarrow\quad \mathcal{C}_H=\eta^{\mu\nu}p_\mu p_\nu+m^2\approx 0
\ee
using Dirac's notion of weak equality $\approx$ for constraints \cite{diracbook}. The naive Hamiltonian $p_\mu\dot{x}^\mu - \mathcal{L}$ vanishes, and the Hamiltonian for the relativistic particle is then $H=\lambda\, \mathcal{C}_H$ where $\lambda$ is a Lagrange multiplier; the Hamiltonian constraint $\mathcal{C}_H$ generates reparametrisations in time.

An alternative way of obtaining the same dynamics is to start from an action that does not have the square root but depends directly on a Lagrange multiplier,
\be
S[x^\mu,\dot{x}^\mu,\lambda] = -\frac{m}{2}\int\dd \tau\left(\frac{1}{\lambda}\eta_{\mu\nu}\dot{x}^\mu\dot{x}^\nu-\lambda\right)\,.
\label{particleaction2}
\ee
For this action to also be invariant under time reparametrisations, the Lagrange multiplier $\lambda$ must transform nontrivially; as $\tau\mapsto\tilde{\tau}=\tilde{\tau}(\tau)$ we must have
\be
\lambda(\tilde\tau)=\lambda(\tau)(\tilde{\tau}'(\tau))^{-1}
\ee
or in other words, the combination $\lambda \;\dd\tau$ must be invariant. $\lambda$ is then analogous to the lapse function in general relativity; it is sometimes referred to as an ``einbein.''

If we treat $\lambda$ as a Lagrange multiplier and do not assign a conjugate momentum to it, the conjugate momenta and Hamiltonian now take the form
\be
p_\mu=-\frac{m}{\lambda}\eta_{\mu\nu}\dot{x}^\nu\,,\qquad H = p_\mu\dot{x}^\mu-\mathcal{L}=-\frac{\lambda}{2m}\left(\eta^{\mu\nu}p_\mu p_\nu+m^2\right)
\ee
and the constraint $\mathcal{C}_H\approx 0$ now arises from the variation with respect to $\lambda$. (An alternative approach would be to view $\lambda$ and its momentum $p_\lambda$ as initially part of the phase space; then we have a primary constraint $p_\lambda\approx 0$ whose conservation in time also leads to $\mathcal{C}_H\approx 0$.)

The two actions can be directly related by observing that the equation of motion for $\lambda$ implies
\be
\lambda = \pm\sqrt{-\eta_{\mu\nu}\dot{x}^\mu\dot{x}^\nu}\,,
\label{lambdaex}
\ee
so that $\lambda\;\dd\tau$ is (up to a sign), as expected, the infinitesimal proper time interval along the particle worldline. (\ref{lambdaex}) can be substituted back into (\ref{particleaction2}); this leads to (\ref{particleaction1}), again up to overall sign. The two actions clearly lead to equivalent Hamiltonian formulations, even though their symmetry structure is different at the Lagrangian level. Indeed, in both cases one obtains a Hamiltonian (first order) action
\be
S[x^\mu,\dot{x}^\mu,p_\mu,\lambda] = \int\dd \tau\left(p_\mu\dot{x}^\mu-\lambda\left(\eta^{\mu\nu}p_\mu p_\nu+m^2\right)\right)\,,
\ee
where if we start from (\ref{particleaction2}) the Lagrange multiplier needs to be redefined as $-\frac{\lambda}{2m}\rightarrow\lambda$.

The two pure connection formulations of main interest in this paper can be seen as analogous to (\ref{particleaction1}) and (\ref{particleaction2}). Also for these theories, we will see that invariance under time reparametrisations is implemented differently in the action, but that the Hamiltonian formulations are equivalent.

\subsection{From Pleba\'{n}ski to a pure connection formulation}
Our presentation of different formulations of general relativity largely follows that of \cite{KrasnovBook}.

We start with the Einstein--Cartan formulation, defined in terms of a real tetrad $E^I$ of 1-forms (frame field) and an $SO(3,1)$ connection with component 1-forms $\omega^I {}_J$.
The tetrad generates a metric $g = \eta_{IJ} E^I \otimes E^J$.
The zero torsion condition $\dee E^I + \omega^I {}_J \wedge E^J = 0$ (Cartan's first structure equation) guarantees that $\omega$ is the Levi-Civita connection for the metric $g$ written in a particular gauge.
This equation and the dynamical Einstein equations in vacuum plus cosmological constant $\Lambda$ can be obtained from the action
\begin{equation}
    \label{Einstein-Cartan action}
    S_\text{EC} \left[ E , \omega \right] = \frac{1}{ 4 \ell_\text{P}^2 } \int \epsilon_{ IJKL } \, E^I \wedge E^J \wedge \left( \Omega^{KL} - \frac{ \Lambda }{ 6 } E^K \wedge E^L \right)\,,
\end{equation}
where $\Omega^{IJ} = \dee \omega^{IJ} + \omega^I {}_K \wedge \omega^{ KJ }$ are the curvature component 2-forms and where capital Latin indices $I,J,K,\ldots$ are raised and lowered by the Minkowski symbol $\eta^{IJ} , \eta_{IJ}$. Here and in the following, we define the Planck length $\ell_\text{P} = \sqrt{ 8 \pi G }$.

Complex formulations then arise from the concept of self-duality. Given a real Lorentzian metric and a volume form, one can define a Hodge star $\star$ mapping 2-forms to 2-forms. The eigenvalues of the Hodge star are $\pm\ii$ so that the space of complex 2-forms decomposes into a sum of self-dual (with eigenvalue $+\ii$) and anti-self-dual (with eigenvalue $-\ii$) subspaces. 
Concretely, we define the Hodge star by $A \wedge \star B = \langle A , B \rangle \varepsilon$ for 2-forms $A,B$, where $\langle A , B \rangle = (1/2) A_{\mu \nu} B_{\rho \sigma} g^{\rho\mu}g^{\sigma\nu}$ for a given metric $g$ and volume form $\varepsilon = \pm \sqrt{-g} \, \dee x^0 \wedge \ldots \wedge \dee x^3$. In index notation, $(\star B)_{\mu \nu} = (1/2) \varepsilon_{\mu \nu} {}^{\rho \sigma} B_{\rho \sigma}$.
In an orthonormal frame $E^I$ with $\varepsilon = E^0 \wedge \ldots \wedge E^3$,  
\be
\star ( E^0 \wedge E^1 ) = - E^2 \wedge E^3\,,\quad\star ( E^2 \wedge E^3 ) = E^0 \wedge E^1\,,
\ee
and so on, so that $\Sigma^i := \ii E^0 \wedge E^i - (1/2) \epsilon^i {}_{jk} E^j \wedge E^k$ are a basis of self-dual 2-forms. 

This basis also arises from another kind of duality on complex bivectors (antisymmetric $\C^4 \otimes \C^4$ tensors), which form a representation of the complexified algebra $\mathfrak{so} (3,1)_\C$.
The map $B^{IJ} \mapsto \ast B^{IJ} = (1/2) \epsilon^{IJ} {}_{KL} B^{KL}$, with the convention $\epsilon^{0ijk} = - \epsilon^{ijk}$, is a linear isomorphism with eigenvalues $\pm \ii$; self-dual bivectors satisfy $\ast B^{IJ} = \ii B^{IJ}$. We can construct a projector onto the self-dual subspace, $P_+ ^{IJ} {}_{KL} := (1/2) \left( \delta^{ [I } _K \delta^{ J] } _L - ( \ii / 2 ) \epsilon^{IJ} {}_{KL}  \right)$, which sends a bivector $B^{IJ}$ to its self-dual part $B_+ ^{IJ}=P_+ ^{IJ} {}_{KL}\,B^{KL}$.
Self-dual bivectors are determined completely by their $0i$ components as they must satisfy $B_+ ^{ij} = - \ii \epsilon^{ij} {}_k B_+ ^{0k}$.
We can hence perform a change of basis on self-dual bivectors so that $B_+ ^{IJ} \to B^i := 2\ii B_+ ^{0i}$.
Then $\Sigma^i = 2\ii \Sigma_+ ^{0i}$ where $\Sigma^{IJ} = E^I \wedge E^J$ is treated as a bivector valued 2-form.
We might call $\Sigma^i$ the {\it self-dual part of the tetrad}.

The self-dual part of the spin connection $\omega^I {}_J$ is recovered by requiring that the projector $P_+$ be compatible with the exterior covariant derivative generated by $\omega^I {}_J$,
\begin{equation}
    2\ii P_+ ^{0i} {}_{IJ} \left( \dee \Sigma^{IJ} + \omega^I {}_K \wedge \Sigma^{KJ} + \omega^J {}_K \wedge \Sigma^{IK} \right) = \dee \Sigma^i + \epsilon ^i {}_{jk} A^j \wedge \Sigma^k \, .
\end{equation}
$A^i$ are the components of the {\it self-dual part of the connection}.
Its curvature $F^i$ turns out to be the self-dual part of the curvature defined from $\omega^I{}_J$, $F^i = 2\ii \Omega_+ ^{0i}$.

One can then complexify the action (\ref{Einstein-Cartan action}) and keep only the self-dual parts of the relevant quantities to construct a diffeomorphism invariant $SO(3,\C)$ gauge theory whose classical dynamics contain the solutions of vacuum Einstein general relativity.
This transformation was first understood by Pleba{\'n}ski \cite{Plebanski} following earlier works such as \cite{cahen1967complex}; see also \cite{PlebanskiIntro,selfdual2} for further details.

We will work with the chiral Pleba{\'n}ski formulation and certain related formulations with fewer independent variables. 
The chiral Pleba{\'n}ski action is given by \cite{KrasnovBook,herfray2016anisotropic}
\begin{equation}
    \label{Plebanski action}
    S [ \Sigma , A , M , \nu ] = \frac{1}{ \ii \, \ell_\text{P} {}^2 } \int \Sigma^i \wedge F_i - \frac{1}{2} M_{ ij } \, \Sigma^i \wedge \Sigma^j + \frac{1}{2} \left( \tr M - \Lambda \right) \nu \,.
\end{equation}
Here $\Sigma^i$ are a triple of complex valued 2-forms, which should be thought of as a single $\mathfrak{so}(3)_\C$ valued 2-form.
Similarly, $A^i$ are the component 1-forms of an $SO(3,\C)$ connection.
The indices $i , j , \ldots$ are with respect to a basis $\tau_i$ on the algebra $\mathfrak{so}(3)_\C$ with structure constants $\epsilon^i {}_{jk}$; these indices are raised and lowered by Kronecker deltas $\delta^{ij} , \delta_{ij}$.
The symmetric matrix
field $M^{ij}$ and the top form $\nu$ generate algebraic constraints on $A^i$ and $\Sigma^i$.
These fields do not have universally agreed upon names in the literature; we will call $M^{ij}$ the {\em auxiliary matrix} and $\nu$ the {\em auxiliary form}.
On-shell, these fields represent the Weyl part of the curvature and a spacetime volume form.
All fields transform under the action of the gauge group $SO(3,\C)$, with infinitesimal transformations 
\be
\delta_\phi A^i = D_A \phi^i\,,\quad \delta_\phi F^i = \epsilon^i {}_{jk} F^j \phi^k\,,\quad \delta_\phi \Sigma^i = \epsilon^i {}_{jk} \Sigma^j \phi^k\,,\quad \delta _\phi M^{ij} = {\epsilon^i}_{kl} M^{kj} \phi^l + {\epsilon^j}_{kl} M^{ik} \phi^l\,.
\ee
The Pleba{\'n}ski action is invariant under these transformations.
The field equations are
\be
D_A \Sigma^i = 0\,,\qquad F^i = M^{ij} \Sigma_j\,,\qquad \Sigma^i \wedge \Sigma^j = \delta^{ij} \nu\,,\qquad \tr M = \Lambda\,.
\label{plebanskieq}
\ee
Here the operator $D_A$ is an exterior covariant derivative which acts on $\mathfrak{so}(3)_\C$ valued functions and forms by $D_A\Sigma^i = \dee \Sigma^i + \epsilon^i {}_{jk} \, A^j \wedge \Sigma^k$ and $D_A \Sigma_i = \dee \Sigma_i - \epsilon^k {}_{ji} A^j \wedge \Sigma_k$, so that raising and lowering of indices commutes with $D_A$.

(\ref{plebanskieq}) are the field equations of complex general relativity; only a small subset of the total solution space corresponds to real Lorentzian solutions that one is usually interested in.
To locate these solutions, {\em reality conditions} must be satisfied on top of the field equations.
They are of
\begin{subequations}
    \label{Plebanski reality condtions first appearance}
    \begin{align}
        & \text{Trace type :}
        \quad
        \text{Re} \left( \Sigma^i \wedge \Sigma_i \right) = 0\,,
        \\[0.5em]
        & \text{Wedge type :}
        \quad
        \Sigma^i \wedge \overline{ \Sigma^j } = 0\,,\quad
        i,j = 1,2,3\,,
    \end{align}
\end{subequations}
where the over-bar denotes complex conjugation.
We can think of these as initial conditions applied to the variables on some initial hypersurface; in general these reality conditions are not immediately compatible with the field equations.
There may be secondary conditions that need to be satisfied to make the total system consistent.

To understand how solutions to the field equations (\ref{plebanskieq}) subject to these reality conditions reproduce solutions of the vacuum Einstein equations (with $\Lambda$) for real Lorentzian metrics, one can show the following (see Chapter 5 of \cite{KrasnovBook}): given a triple of complex 2-forms $\Sigma^i$ satisfying $\Sigma^i \wedge \overline{ \Sigma^j } = 0$ and $\Sigma^i \wedge \Sigma^j = - 2\ii \delta^{ij} \varepsilon_{\Sigma}$ for some real top-form $\varepsilon_\Sigma$, there exists a tetrad $E^I$ of real 1-forms that allow us to write $\Sigma^i$ either as
\begin{subequations}
    \begin{align}
        \label{basis of self dual forms}
        & \Sigma^1 = \ii E^0 \wedge E^1 - E^2 \wedge E^3
        \,, \quad
        \Sigma^2 = \ii E^0 \wedge E^2 - E^3 \wedge E^1
        \,, \quad
        \Sigma^3 = \sigma \left( \ii E^0 \wedge E^3 - E^1 \wedge E^2 \right)\,,\quad{\rm or}
        \\[0.5em]
        \label{basis of anti self dual forms}
        & \Sigma^1 = E^0 \wedge E^1 - \ii E^2 \wedge E^3
        \,, \quad
        \Sigma^2 = E^0 \wedge E^2 - \ii E^3 \wedge E^1
        \,, \quad
        \Sigma^3 = \sigma \left( E^0 \wedge E^3 - \ii E^1 \wedge E^2 \right)\,,
    \end{align}
\end{subequations}
where $\sigma = \pm 1$. For self-consistency we must then have $\varepsilon_\Sigma = E^0 \wedge \ldots \wedge E^3$.
In the first case (\ref{basis of self dual forms}), the $\Sigma^i$ constitute a basis of the self-dual 2-forms with respect to the Hodge star operator defined from the metric $g = \eta_{IJ} E^I \otimes E^J$ in the orientation set by $\varepsilon_\Sigma$, where $\eta_{IJ}$ may have either of the signatures $(- + + +)$ or $(+ - - -)$. In the second case (\ref{basis of anti self dual forms})  the $\Sigma^i$ constitute a basis of the anti-self-dual forms with respect to the same $g$ in the same orientation.

Alternatively, if the reality conditions (\ref{Plebanski reality condtions first appearance}) are  satisfied, one may directly construct the metric $g_\Sigma$ and volume form $\varepsilon_\Sigma$ using the Urbantke formula \cite{selfdual2, urbantke}
\be
\varepsilon_\Sigma = \frac{ \ii }{6} \Sigma ^i \wedge \Sigma_i\,,\qquad g_\Sigma \left( \xi , \eta \right) \, \varepsilon_\Sigma = -\frac{ \ii }{6} \epsilon_{ijk} \, i_\xi \Sigma^i \wedge i_\eta \Sigma^j \wedge \Sigma^k\,.
\label{urbantkeformula}
\ee
Here $\xi$ and $\eta$ are arbitrary vector fields and $i_\xi\Sigma^i:=\Sigma^i(\xi,\cdot)$ is the insertion of a vector into the first slot of the 2-form $\Sigma^i$. In the case where the $\Sigma^i$ may be decomposed as in (\ref{basis of self dual forms}), the Urbantke metric is $g_\Sigma = \sigma\eta_{IJ} E^I \otimes E^J$ where $\eta$ has signature $(-+++)$. On the other hand, when the $\Sigma^i$ decompose as in (\ref{basis of anti self dual forms}) the Urbantke metric is $g_\Sigma = \ii \sigma\left( \eta_{IJ} E^I \otimes E^J \right)$.
In this case, one could construct a real Lorentzian metric $g_{ \widetilde \Sigma }$ from the rescaled 2-forms $\widetilde \Sigma^i := - \ii \Sigma^i$.

The details of the correspondence of (\ref{plebanskieq}) with general relativity depend on which of the cases (\ref{basis of self dual forms})--(\ref{basis of anti self dual forms}) one is in. For the most straightforward case in which $\Sigma^i$ decompose as in (\ref{basis of self dual forms}) with $\sigma=+1$, this can be found in \cite{PlebanskiIntro}.
In short, the field equation $D_A \Sigma^i = 0$ is the chiral zero torsion condition, which tells us that the connection $A^i$ is the self-dual part of the Levi-Civita connection associated to $\Sigma^i$: $A^i =\ii \omega^{0i} - (1/2) \epsilon^i {}_{jk} \, \omega^{jk}$ where $\omega$ is the unique $SO(3,1)$ connection satisfying $\dee E^I + \omega^I {}_J \wedge E^J = 0$, with $E^I$ the tetrad derived from $\Sigma^i$.
Then the field equation $F^i = M^{ij} \Sigma_j$ tells us that the curvature forms $F^i$ are self-dual with respect to the metric $g_\Sigma$.
This is equivalent to the trace-free Einstein condition $R_{\mu \nu} - \frac{1}{4} R g_{\mu \nu} = 0$.
Then the final field equation $\tr M = \Lambda$ gives the missing equation $R_{\mu\nu} = \Lambda g_{\mu\nu}$.\footnote{In the unimodular version of general relativity, this last relation is not imposed as a field equation, one only has the weaker statement $\nabla_\mu R=0$ and $\Lambda$ is seen as a integration constant. Similarly, in the Pleba\'{n}ski formulation the first two equations of (\ref{plebanskieq}) would already imply $D_A M^{ij}=0$ and so $\tr M={\rm const}$. We are not aware of a ``unimodular'' Lagrangian formulation that reproduces the first three equations in (\ref{plebanskieq}) but not the last one.}
Hence, the trace-free part of $M^{ij}$ corresponds to the Weyl curvature while the trace part corresponds to the Ricci curvature.

The other cases can be mapped to this first case by redefining the 2-forms $\Sigma^i$ by constant factors. For instance, if $\Sigma^i$ satisfy (\ref{basis of self dual forms}) but with $\sigma=-1$, one can define $(\Sigma')^i:=-\Sigma^i$ such that $(\Sigma')^i$ fall into the case we just discussed. If the $\Sigma^i$ solve the equations (\ref{plebanskieq}) for some $A^i$, $M^{ij}$ and a particular $\Lambda$, then $(\Sigma')^i$ solve (\ref{plebanskieq}) for the same $A^i$, $(M')^{ij}:=-M^{ij}$ and consequently for the opposite sign of $\Lambda$. The Urbantke metric $g_{\Sigma'}$ hence satisfies $R_{\mu\nu}=-\Lambda g_{\mu\nu}$, but then the Urbantke metric $g_\Sigma$ for the original $\Sigma^i$ satisfies $R_{\mu\nu}=\Lambda g_{\mu\nu}$ just as in the first case. This argument can be extended to the other cases to show that the Urbantke metric always has $R_{\mu\nu}=\Lambda g_{\mu\nu}$, for all of the four cases. This implies that, if we assume $\Lambda$ to be real, the cases in (\ref{basis of anti self dual forms}) have no solutions in (\ref{plebanskieq}): such solutions would require an imaginary Urbantke metric but real Ricci tensor. We will see this explicitly in the cosmological models we are studying later on.\\

The idea is now, starting from the action (\ref{Plebanski action}), to ``integrate out'' fields and construct a new action with fewer independent variables. The first step leads to a {\em first order chiral connection action}, which does not appear frequently in the literature but is discussed in detail in \cite{KrasnovBook} and to some extent in \cite{herfray2016anisotropic}. It also appears as an intermediate step in the construction of \cite{purespinc}, and has been dubbed the ``instanton representation of Plebanski gravity'' in \cite{eyoita}.
To get the first order action, we solve the field equation $F^i = M^{ij} \Sigma_j$ for $\Sigma^i$ and insert the result, $\Sigma^i=( M^{-1} )^{ij} F_j$, back into the action.
The resulting action reads
\begin{equation}
    \label{Chiral first order action}
    S_\text{FO} [ A , M , \nu ] = \frac{1}{ 2 \ii \, \ell_\text{P}^2 } \int M^{-1}_{ ij } F^i \wedge F^j + \left( \tr M - \Lambda \right) \nu
\end{equation}
To check that this off-shell substitution does not change the dynamics of the theory, denote $\Sigma_F ^i := \left( M^{-1} \right)^{ij} F_j$.
Then the field equations for the first order theory read
\be
\label{First order field equations}
D_A \Sigma_F ^i = 0\,,\qquad \Sigma_F ^i \wedge \Sigma_F ^j = \delta^{ij} \nu\,,\qquad \tr M = \Lambda\,.
\ee
By inspection, we see that these field equations are exactly the field equations of the chiral Pleba{\'n}ski theory under the substitution $\Sigma^i \to \Sigma_F ^i$.
Hence the constraint $F^i = M^{ij} \Sigma_j$ commutes with the variational principle.
Using the Bianchi identity $D_A F^i = 0$, the first field equation $D_A \Sigma_F ^i = 0$ implies $\left( D_A M^{-1}_{ij} \right) F^j = 0$ which has only first order derivatives of $M$ and $A$; hence this is a first order theory.

The reality conditions for the first order theory are the reality conditions for the generalised Pleba{\'n}ski theory where $\Sigma^i$ is replaced with $\Sigma_F ^i$,
\begin{subequations}
    \label{realitycond}
    \begin{align}
        & \text{Trace type :}
        \quad
        \text{Re} \left( \Sigma_F ^i \wedge \Sigma_{F \, i} \right) = 0\,,
        \\
        & \text{Wedge type :}
        \quad
        F^i \wedge \overline{F^j} = 0\,,
        \qquad
        i,j = 1,2,3\,.
    \end{align}
\end{subequations}
Notice that these conditions now contain first derivatives of the connection whereas previously the reality conditions were algebraic conditions on the Pleba{\'n}ski 2-forms which could be imposed on some initial surface without any knowledge of the dynamics. Now, we should see the reality conditions as further conditions applied on top of the field equations. We must then check that these conditions are compatible with the dynamics and do not produce only trivial solutions. We explore this issue in greater detail for spatial homogeneous models later on.

The final step in this process leads to the {\em chiral pure connection action} first introduced in \cite{KrasnovPureConn} and further explored in \cite{KrasnovGauge}.
Here one eliminates the remaining auxiliary fields $M^{ij}$ and $\nu$ from the first order action by solving the equations $\Sigma_F ^i \wedge \Sigma_F ^j = \delta^{ij} \nu$ for $M$ in terms of $F$ and $\nu$, and then solving $\tr M = \Lambda$ for $\nu$ in terms of $F$, yielding a theory in which the only field is a connection.
The resulting action reads
\begin{equation}
    \label{Pure connection action}
    S_{{\rm PC}} [ A ] = \frac{1}{ \ell_\text{P}^2\, \Lambda } \int \left( \tr \sqrt{ X } \right)^2 \varepsilon_X
\end{equation}
where $\varepsilon_X$ is a fixed, nowhere vanishing top-form and a matrix-valued function $X^{ij}$ is defined implicitly by $F^i \wedge F^j = 2\ii \, X^{ij} \varepsilon_X$.
In the models that we will examine, this will always take diagonal form $X^{ij} = \chi^i \, \delta^{ij}$; in this case computing the square root matrix is straightforward as $( \sqrt{X} )^{ij} = \sqrt{ \chi^i } \, \delta^{ij}$ where we must choose a branch of the complex square root function. In the more general case one would need to provide a prescription for how to define a matrix square root.

Variations of the action yield a single field equation,
\be
  D_A\left(\frac{ \tr \sqrt{X} }{ \Lambda } \left( X^{-1/2} \right)^{ij} F_j\right)=0
\ee
where $X^{-1/2} = ( \sqrt{X} ) ^{-1}$ is the inverse of the matrix square root. The expression inside the brackets now corresponds to the Pleba{\'n}ski 2-form $\Sigma_X^i$, seen as a function of $F^i$, and one can check that it satisfies $\Sigma_X^i\wedge\Sigma_X^j\sim\delta^{ij}$ by definition of $X^{ij}$. Finally, $M_X:=\sqrt{X}\frac{\Lambda}{\tr \sqrt{X}}$ evidently satisfies $\tr M_X = \Lambda$. Hence this theory still encodes the full dynamics of the original Pleba{\'n}ski action.

The reality conditions now take the form
\begin{subequations}
    \begin{align}
        & \text{Trace type :}
        \quad
        \text{Re} \left( \tr \sqrt{X} \right) = 0\,,
        \\[0.5em]
        & \text{Wedge type :}
        \quad
        F^i \wedge \overline{F^j} = 0\,,
        \qquad 
        i,j = 1,2,3\,.
    \end{align}
\end{subequations}
These are extremely complicated when seen as conditions on the connection $A^i$, which is a significant drawback of this formalism when applied to Lorentzian general relativity.

\section{Spatially homogeneous models}
\label{Homogeneity section}

We assume a spacetime topology $\mathcal{M} = \R \times {\bf S}$ where ${\bf S}$ is a simply connected 3-manifold.
It is then natural to restrict to coordinate charts that preserve the product structure of $\mathcal{M}$, so that the only acceptable coordinate transformations are of the form $(t,x^a) \to ( t'(t) ,  x' {}^a (x^b) )$.
Given some coordinate chart $\left( t , x^a \right)$ the component 1-forms of a connection can be written as
\begin{equation}
    \label{Foliated Connection}
    A^i = \varphi^i \, \dee t + A^i _a \, \dee x^a\,.
\end{equation}

For cosmological Bianchi models, we further restrict to the case where ${\bf S}$ is a connected 3-dimensional Lie group, so that for each $s \in {\bf S}$ there is a left translation map $\ell_s : {\bf S} \to {\bf S} , r \mapsto sr$ which is a diffeomorphism.
The requirement of {\em spatial homogeneity} means that our connection should be invariant under the pull-back of these left translation maps.
To construct homogeneous connections, one can define a globally independent, non-vanishing triad of 1-forms $e^{\bf a}$ that are invariant under the pull-back of these left translation maps, called a {\em Cartan frame}. Such a frame satisfies the Maurer--Cartan relations
\begin{equation}
    \dee e^{\bf a} + \frac{1}{2} f ^{ \bf a } {}_{\bf bc} \, e^{\bf b} \wedge e^{\bf c} = 0\,,
\end{equation}
where $f ^{\bf a} {}_{\bf bc}$ are the structure constants of the Lie algebra of ${\bf S}$ in some basis $\mathcal{J}_{\bf a}$.
We can use a Cartan frame to construct a homogeneous connection $A^i = \varphi^i (t) \, \dee t + A^i _{\bf a} (t) \, e^{\bf a}$.

General gauge transformations are generated by group valued functions $U \in \mathcal{M} \to SO(3,\C)$ through the usual
mechanism\footnote{For infinitesimal transformations
$U = \boldsymbol 1 + \phi^i \tau_i$, this of course reduces to our previous expression $\delta_\phi A = D_A\phi$.}
\begin{equation}
    A
    \; \longrightarrow \;
    A^U = U^{-1} A \, U + U^{-1} \dee U\,.
\end{equation}
A homogeneous gauge transformation is generated by a group valued function $U(t)$ which has only time dependence.
We can expand such a gauge transformation as
\begin{equation}
    A^U = U^{-1} \left( \varphi U + \dot U \right) \dee t + U^{-1} A_{\bf a} \, U e^{\bf a}
\end{equation}
where here and in the following $\dot{}$ denotes differentiation with respect to $t$.
Now notice that the differential equation $\dot U = - \varphi \, U$ is solved by the path ordered exponential
\begin{equation}
    U (t) = \mathcal{P} \exp \left( - \int_0 ^t \dee \tau \, \varphi (\tau) \right) U_0
\end{equation}
for arbitrary $U_0 \in SO(3,\C)$.
Hence, it is always possible to bring a spatially homogeneous connection into the form $A^i = A^i _{\bf a} \, e^{\bf a}$ under the action of homogeneous gauge transformations. We will assume this form in the following, and not include the spurious part $\varphi^i\,\dee t$.

One of our goals is to understand the dynamics of chiral connection theories for spatially homogeneous models.
To do this, we would like to apply a symmetry reduction to the action functional by constructing ans\"atze for our fields $A^i (z) , M^{ij} (z) , \nu (z)$ in terms of new variables $z^A ( t , {\boldsymbol x} )$ that satisfy the symmetry requirements, and substituting these ans\"atze into, e.g., the first order action (\ref{Chiral first order action}) to get a new action $\widetilde S \left[ z \right]$ over the reduced variables $z^A$.

The {\it symmetric criticality principle} \cite{palais1979principle} implies that, e.g., for the case ${\bf S} = S^3 \cong SU(2)$ of main interest in this paper, the compactness of the group guarantees that the restriction to $SU(2)$ translation-invariant connections commutes with the variational principle.
We also wish to make a further restriction to {\it diagonal connections} which will be defined in the next section.
This restriction does not appear to be due to invariance under some group of transformations, so the symmetric criticality principle does not apply in an obvious way.
We can check by hand that our restrictions commute with the variational principle, i.e., that the field equations that we obtain from the reduced action $\widetilde S \left[ z \right]$ are equivalent to equations that we get when we substitute our ans\"atze into the general field equations (\ref{First order field equations}).
Schematically, we must confirm that the following diagram commutes:
\begin{equation}
    \label{commuting diagram}
    \begin{array}{ccc}
        \text{General Action} & \overset{ \text{Variations} }{ \xrightarrow{ \hspace*{3cm} } } & \text{Field Equations}  \\
        \\
        \bigg \downarrow & \left( \text{Restriction of fields} \right) & \bigg \downarrow \\
        \\
        \text{Reduced Action} & \overset{ \text{Variations} }{ \xrightarrow{ \hspace*{3cm} } } & \text{Reduced Field Equations}
    \end{array}
\end{equation}
It would be an illuminating exercise to substitute the most general ansatz for a spatially homogeneous connection $A^i = \varphi^i (t) \dee t + A^i _{\bf a} (t) e^{\bf a}$ into the first order action to see what conditions the commutativity of the diagram (\ref{commuting diagram}) puts on the structure constants $f^{\bf a} {}_{\bf bc}$.
A similar calculation was done in the metric formalism in \cite{torre1999midi}.

\section{Diagonal Bianchi IX model}
\label{Diagonal symmetric models section}

\newcommand{\jj}{ \mathcal{J} }
\newcommand{\cc}{ \mathcal{C} }

We are now ready to start the main part of our investigation, which is to study the cosmological Bianchi IX model in the context of the chiral first order theory defined by (\ref{Chiral first order action}). The Bianchi IX model is discussed in section 6.7 of \cite{KrasnovBook} and our analysis will initially follow the presentation there. We will then discuss additional solutions to the reality conditions (\ref{realitycond}) beyond the physically most relevant ones appearing in \cite{KrasnovBook}. We will also construct a Hamiltonian formalism in which the reality conditions and their associated consistency conditions can be viewed as (second class) constraints in Dirac's formalism for constrained Hamiltonian systems.

In the Bianchi IX model spacetime is taken to be of the form $\mathcal{M} = \R \times S^3$ and we require the fields to be spatially homogeneous, as described in section \ref{Homogeneity section}.
A further simplification which is commonly made in the literature is that the connection is assumed to be diagonal, which means that there exists a combination of a homogeneous gauge transformation and a change of frame that brings the connection into the form $A^i = C^i (t) \, \delta^i _{\bf a} \, e^{\bf a}$ (no sum over $i$).
Diagonal connection models are discussed in the context of loop quantum cosmology in \cite{bojowaldhomo}.
In general, the requirement that the connection be diagonal is a restriction on the variables, and we will need to confirm that this restriction commutes with the variational principle.

We can normalise a Cartan frame $e^{\bf a}$ so that it satisfies $\dee e^{\bf a} + \sqrt{k} \, \epsilon^{\bf a} {}_{{\bf b}{\bf c}} \, e^{\bf b} \wedge e^{\bf c} = 0$ where $k$ is a positive constant; this $k$ corresponds to the spatial curvature parameter commonly used in cosmology. Then $\varepsilon_e := e^1 \wedge e^2 \wedge e^3$ defines a volume form on $S^3$. We will assume that our cosmological model describes a compact and connected 3-manifold $\mathcal{V} \subseteq S^3$ (which can be, and is usually taken to be, all of $S^3$). Then we can define a ``fiducial'' 3-volume $V := \int_\mathcal{V} \varepsilon_e$, which is invariant under $SU(2)$ translations.\footnote{This fiducial volume is commonly used in spatially flat models to make the action finite, see, e.g., \cite{fiducial} for a recent discussion. In principle one could choose $\mathcal{V}$ to be a proper submanifold of $S^3$ also in the Bianchi IX case.}

We now want to construct a reduced action as described in the previous section \ref{Homogeneity section}.
The reduced variables are complex scalars $z^A \left( t , { \boldsymbol x } \right) = \{C^i (t) , M^i (t) , \rho(t)\}$ with $i = 1,2,3$.
Our homogeneous and diagonal connection ansatz, and the resulting expression for the curvature, are
\begin{equation}
    \label{Connection and curvature ansatz}
    A^i = \ii V ^{-1/3} C^i \, e^i\,,
    \qquad
    F^1 = \ii V ^{-1/3} \dot C^1 \, \dee t \wedge e^1 - V ^{-2/3} \left( \ii \kappa C^1 + C^2 C^3 \right) \, e^2 \wedge e^3
    \quad \text{etc.}\,,
\end{equation}
where $\kappa := 2 V ^{1/3} \sqrt{k}$ is a convenient shorthand.
Then the top-form matrix $F^i \wedge F^j$ is diagonal with non-vanishing entries
\begin{equation}
    F^1 \wedge F^1 = 2 \ii V^{-1} \dot C^1 \left( \ii \kappa C^1 + C^2 C^3 \right) \varepsilon_e \wedge \dee t
    \quad \text{etc.}
\end{equation}
In the first order action (\ref{Chiral first order action}) the top-form matrix $F^i \wedge F^j$ contracts fully with the inverse auxiliary matrix $M^{-1} _{ij}$; since all of the off-diagonal entries of $M^{-1} _{ij}$ vanish in this contraction, we can take $M^{ij}$ to be diagonal.
We parametrise $M^{ij} = \ell_\text{P} ^{-2} \, M^i \, \delta^{ij}$ where the factor $\ell_\text{P} ^{-2}$ sets the units of $M^{ij}$ to $\left( \text{length} \right)^{-2}$ which is required to make the action dimensionless; in our conventions, coordinates and the connection and curvature forms are all dimensionless.
The auxiliary form is parametrised as $\nu = 2 \ii\, V ^{-1} \ell_\text{P}^4\, \rho \, \varepsilon_e \wedge \dee t$ (powers of $V$ are always chosen to make $V$ drop out of the dynamical formalism).
We substitute these forms of $F^i$, $M^{ij}$ and $\nu$ into the first order action, which yields an expression of the form $\left( \int_\mathcal{V} \varepsilon_e \right) \int \dee t \, L \left( C , M , \rho , \dot C \right)$. The spatial integral gives a factor $V$ and we get
\begin{equation}
    \label{Bianchi IX action general}
    \widetilde S \left[ C , M , \rho \right] = \int \dee t\left[\left( \frac{ \ii \kappa C^1 + C^2 C^3 }{ M^1 } \, \dot C^1 + \text{permutations} \right) - \rho \left( \ell_\text{P}^2 \Lambda - { \textstyle \sum }_{i=1}^3 \, M^i \right) \right]\,,
\end{equation}
which is independent of the fiducial volume $V$ as desired.

In our parametrisation of $\nu$ in terms of $\rho$, we used a particular choice of time coordinate $t$. In order to maintain invariance of $\widetilde S$ under reparametrisations of time, $\rho$ must transform as
\begin{equation}
    \label{scalar density transformation law}
    \rho (t) = \frac{ \dee t' }{ \dee t } \rho' (t')
\end{equation}
or in other words, under a time reparametrisation we must have $\rho\,\dee t = \rho'\,\dee t'$. $\rho$ is then again analogous to a lapse function in general relativity. 

To bring the action into a more convenient form, we can introduce a new triple of complex scalars $P_i (t)$ to replace $M^i (t)$ using the variable redefinitions
\begin{equation}
    \label{Inertial matrix ansatz}
    M_1 = \frac{ \ii \kappa C^1 + C^2 C^3 }{ P_1 }
    \quad \text{etc.}
\end{equation}
In terms of these new variables, the reduced action becomes
\begin{equation}
    \label{Bianchi IX Action Hamiltonian Form}
    S \left[ C , P , \rho \right] = \int \dee t\left[P_i \, \dot C^i - \rho \left( \ell_\text{P}^2 \Lambda - \left( \frac{ C^2 C^3 }{ P_1 } + \text{permutations} \right) - \ii \kappa \sum_{i=1} ^3 \frac{ C^i }{ P_i } \right) \right]\,.
\end{equation}
This action has some similarities with the action (\ref{particleaction2}) for a relativistic particle; in particular we have a lapse-like variable $\rho$ similar to the ``einbein'' $\lambda$ used there. However, (\ref{Bianchi IX Action Hamiltonian Form}) is already in Hamiltonian form, with $C^i$ and $P^i$ analogous to position and momentum variables for a relativistic particle, in contrast with (\ref{particleaction2}) defined in terms of position variables only.

As discussed in section \ref{Homogeneity section} we now need to check that the dynamics that arise from this reduced action coincide with the dynamics of the general chiral first order theory with our symmetry reductions applied on-shell by hand.

We substitute our ans\"atze for $A^i$ and the other fields into the first order field equations (\ref{First order field equations}) and rearrange to find equations of motion in terms of the reduced variables.
We begin by constructing 
\begin{equation}
    \label{Sigma_F expression}
    \Sigma_F ^i = \left( M^{-1} \right)^{ij} F_j = V^{-2/3} \ell_\text{P}^2 \left( \frac{ \ii V^{1/3} P_1 \dot C^1 }{ \ii \kappa C^1 + C^2 C^3 } \, \dee t \wedge e^1 - P_1 \, e^2 \wedge e^3 \right)
    \quad \text{etc.}
\end{equation}
Then we examine the condition $\Sigma_F ^i \wedge \Sigma_F ^j = \delta^{ij} \nu$, which reduces to three equations
\begin{equation}
    \Sigma_F ^1 \wedge \Sigma_F ^1 = \frac{ 2 \ii V^{-1}\, \ell_\text{P}^4\,  (P_1)^2 \, \dot C^1 }{ \ii \kappa C^1 + C^2 C^3 } \, \varepsilon_e \wedge \dee t
    =
    2\ii V^{-1} \ell_\text{P}^4\, \rho \, \varepsilon_e \wedge \dee t
    \quad \text{etc.}
\end{equation}
These yield three first order equations of motion
\begin{equation}
    \dot C^1 = \rho \, \frac{ \ii \kappa C^1 + C^2 C^3 }{ (P_1)^2 }
    \quad
    \text{etc.}
    \label{eqmo1}
\end{equation}
We substitute these back into (\ref{Sigma_F expression}) to get simpler expressions for $\Sigma_F ^i$\,,
\begin{equation}
    \Sigma_F ^1 = V^{-2/3} \ell_\text{P}^2 \left( \frac{ \ii V^{1/3} \rho }{ P_1 } \, \dee t \wedge e^1 - P_1 \, e^2 \wedge e^3 \right)\quad \text{etc.}\,,
    \label{simplesigmaf}
\end{equation}
and
\be
D_A \Sigma_F ^1 = \dee \Sigma_F ^1 + A^2 \wedge \Sigma_F ^3 - A^3 \wedge \Sigma_F ^2 = V^{-2/3} \ell_\text{P}^2 \,\rho\left( \frac{ \ii \kappa }{ P_1 } + \frac{ C^3 }{ P_2 } + \frac{ C^2 }{ P_3 } - \frac{ \dot P_1 }{ \rho } \right) \dee t \wedge e^2 \wedge e^3\,.
\ee
Hence, from $D_A\Sigma_F^i=0$ we get  three more first order equations of motion
\begin{equation}
    \dot P_1 = \rho \left( \frac{ \ii \kappa }{ P_1 } + \frac{ C^3 }{ P_2 } + \frac{ C^2 }{ P_3 } \right)
    \quad \text{etc.}
    \label{eqmo2}
\end{equation}
Finally we have the constraint $\tr M = \Lambda$ which becomes
\begin{equation}
    \ell_\text{P}^2 \Lambda - \left( \frac{ C^2 C^3 }{ P_1 } + \text{permutations} \right) - \ii \kappa \sum_{i=1}^3 \frac{ C^i }{ P_i } = 0\,.
    \label{eqmo3}
    \end{equation}
$\rho$ does not have an equation of motion and can be chosen arbitrarily, consistent with the fact that it appears as a Lagrange multiplier related to the time reparametrisation freedom in the model.

Now we compute the Euler--Lagrange equations of motion for the reduced action (\ref{Bianchi IX Action Hamiltonian Form}) and compare them with (\ref{eqmo1}), (\ref{eqmo2}) and (\ref{eqmo3}) to confirm that the symmetry reductions commute with the variational principle.
We have an action of the form $S = \int \dee t\left(P_i \dot C^i - \rho H\right)$ where $H$ is a holomorphic function of the configuration variables $C^i , P_i$ but not their derivatives.
The Euler--Lagrange equations are given by
\begin{equation}
    \label{Lagrange-Hamilton equations 1}
    \dot C^i = \rho \, \frac{ \partial H }{ \partial P_i }\,,\qquad
    \dot P_i = - \rho \, \frac{ \partial H }{ \partial C^i }\,.
\end{equation}
Computing the derivatives explicitly yields
\be
\frac{ \partial H }{ \partial C^1 } = - \frac{ \ii \kappa }{ P_1 } - \frac{ C^3 }{ P_2 } - \frac{ C^2 }{ P_3 }
        \quad \text{etc.}\,,\quad
 \frac{ \partial H }{ \partial P_1 } = \frac{ \ii \kappa C^1 + C^2 C^3 }{ (P_1)^2 }
        \quad \text{etc.}      
\ee
Finally, variations with respect to the $\rho$ variable gives the constraint
\begin{equation}
    H = \Lambda _\text{P} - \left( \frac{ C^2 C^3 }{ P_1 } + \text{permutations} \right) - \ii \kappa \sum_{i = 1}^3 \frac{ C^i }{ P_i } = 0
\end{equation}
where $\Lambda _\text{P} := \ell_\text{P}^2 \Lambda$ is the dimensionless expression of $\Lambda$ in Planck units.
These equations of motion are equivalent to (\ref{eqmo1}), (\ref{eqmo2}) and (\ref{eqmo3}): symmetry reduction commutes with the variational principle.

\subsection{Holomorphic Hamiltonian system}

The Lagrangian is a manifestly complex valued function of complex valued inputs, and so the usual notion of Legendre transform that maps from a real Lagrangian to the corresponding Hamiltonian theory does not apply. 
However, the Lagrangian is polynomial in the variables and their derivatives and has no dependence on the complex conjugate variables, and we can construct a {\it holomorphic} Hamiltonian system with identical dynamics by inspection.
The phase space is $\mathcal{P} = \C^6$ with complex Cartesian coordinates $\left( C^i , P_i \right)$ and the symplectic form is a holomorphic 2-form $\Omega := \dee P_i \wedge \dee C^i$ that generates a Poisson bracket on the function
algebra\footnote{
The algebra of functions $f : \mathcal{P} \to \C$ satisfying $\left( \partial f / \partial \, \overline{ w^n } \right) = 0$ where $w^n$ denotes the complex Cartesian coordinates in general and $\overline{ w^n }$ denotes their complex conjugates.
}
$C_\text{hol} \left( \mathcal{P} \right)$ given by
\begin{equation}
    \left\{ A , B \right\} = \pd{ A }{ C^i } \pd{ B }{ P_i } - \pd{ B }{ C^i } \pd{ A }{ P_i }\,;
\end{equation}
in particular $\left\{ C^i , P_j \right\} = \delta^i _j$.
The surface of physical states $\mathcal{S} \subset \mathcal{P}$ is the zero locus of the phase space function $H$.
The time evolution of phase space functions $f \in C_\text{Hol} ( \mathcal{P} )$ is governed by 
\begin{equation}
    \dot f = \rho \left\{ f , H \right\}
\end{equation}
where $\rho \in C_\text{Hol} ( \mathcal{P} )$ is an arbitrary function, a Lagrange multiplier encoding the time reparametrisation freedom.
The time evolution is consistent with the restriction to $\mathcal{S}$ since $\left\{ H , H \right\} = 0$;
physical states evolve into physical states.
Computing $\dot C^i = \rho \left\{ C^i , H \right\}$ and $\dot P_i = \rho \left\{ P_i , H \right\}$ yields (\ref{Lagrange-Hamilton equations 1}), and the dynamics of this Hamiltonian theory are identical to those of the Lagrangian theory.

This holomorphic approach will be insufficient when it comes to applying the reality conditions.
The wedge reality condition $F^i \wedge \overline{ F^j } = 0$ has explicit dependence of the complex conjugate variables.
The three constraints that they generate on the variables will not be holomorphic.

\subsection{Metric reconstruction}
\label{reconstruct metric}

Even though the first order connection formulation does not need a spacetime metric for its definition, we discussed how one can recover a notion of metric using the Urbantke formula (\ref{urbantkeformula}). This object is particularly useful when trying to give an interpretation to different branches of solutions to the reality condition, as we will do in the following. 

To reconstruct a spacetime metric from (\ref{urbantkeformula}), we need to use the derived Pleba\'nski two-forms for our model, defined in (\ref{simplesigmaf}). From these we construct a volume form via
\begin{equation}
    \varepsilon_\Sigma = \frac{ \ii }{ 6 } \Sigma^i \wedge \Sigma_i = V^{-1} \ell_\text{P}^4\, \rho \, \dee t \wedge \varepsilon_e 
\end{equation}
which defines the Urbantke metric implicitly through
\begin{equation}
    g_\Sigma \left( \xi , \eta \right) \, \varepsilon_\Sigma =- \frac{ \ii }{ 6 } \epsilon_{ijk} \, i_\xi \Sigma^i \wedge i_\eta \Sigma^j \wedge \Sigma^k\,.
\end{equation}
We compute the Urbantke metric for our system to be 
\begin{equation}
\label{Urbantke metric}
    g_\Sigma = - \frac{ \ell_\text{P}^2\, \rho^2 }{ P_1 P_2 P_3 } \, \dee t \otimes \dee t + V ^{-2/3} \ell_\text{P}^2 \, P_1 P_2 P_3 \, \sum_{i=1}^3 \frac{1}{ (P_i)^2 } \, e^i \otimes e^i\,.
\end{equation}
The most interesting case for cosmological applications is when both $\rho$ and all $P_i$ are real. Indeed, in this case we obtain a Lorentzian cosmological metric of signature $(-+++)$ or $(+---)$ depending on the sign of the product $P_1 P_2 P_3$, where $\dee t$ denotes the timelike direction as expected. Moreover, the Lagrange multiplier $\rho$ is then related to the usual lapse function of general relativity by $N = \ell_\text{P} |P_1 P_2 P_3|^{-1/2} \rho$. In general, we see that $P_i$ encode metric degrees of freedom, as one would have expected given that they appear as canonically conjugate to the connection $C^i$.

An orthonormal tetrad of one-forms $E^I$, such that $g_\Sigma = \eta_{ IJ } E^I \otimes E^J$ with $\eta_{IJ}$ having signature $(- + + +)$, can be defined by
\begin{equation}
    E^0 = \frac{ \ell_\text{P} \, \rho }{ \sqrt{P_1 P_2 P_3}}\, \dd t\,,
    \quad , \quad
    E^i = V^{-1/3} \frac{ \ell_\text{P} \sqrt{ P_1 P_2 P_3 }}{ P_i } \, e^i\,.
\end{equation}
Just as the Urbantke metric, this tetrad will in general be complex-valued. One also needs to set a convention for the square root of a complex variable. In the case where $\rho$ and all $P_i$ are real with $P_1 P_2 P_3 < 0$, this tetrad would be purely imaginary even though the Urbantke metric is real Lorentzian. One could then alternatively define a real tetrad using the opposite sign conventions for $\eta_{IJ}$, as discussed below (\ref{urbantkeformula}).

\subsection{Obtaining Lorentzian solutions}
\label{Lorentzian Bianchi}

In order to obtain solutions representing those of Lorentzian general relativity, we need to apply the reality conditions (\ref{realitycond}). We begin with the trace condition $\text{Re} \left( \Sigma^i _F \wedge \Sigma_{ F \, i } \right) = 0$, with $\Sigma_F^i$ given in (\ref{simplesigmaf}). It is immediate to see that the trace condition becomes the statement that $\text{Im} \left( \rho \right) = 0$, i.e., $\rho$ needs to be a real function. We can impose this condition without any implications for any dynamical equations, given that $\rho$ is a Lagrange multiplier not constrained by the dynamics.

Then, given the ansatz (\ref{Connection and curvature ansatz}), the wedge reality conditions $F^i \wedge \overline{ F^j } = 0$ become 
\be
\text{Im} \left( \overline{ \left( \ii \kappa C^1 + C^2 C^3  \right)} \, \dot C^1 \right) = 0\quad\text{etc.}
\ee
We can use the dynamical equations (\ref{eqmo1}) to rewrite these as $\text{Im} \left( P_i {}^2 \right) = 0$.
Hence, each of the $P_i$ variables must take only real or pure imaginary values, and there are four distinct solution branches (up to index permutations) depending on how many of the $P_i$ are real and imaginary.
The most straightforward interpretation of the four cases is in terms of the resulting Urbantke metric (\ref{Urbantke metric}). The case where all $P_i$ are real will produce a real-valued metric with Lorentzian signature and with the time direction as the negative eigenvalue direction, i.e., a conventional cosmological metric. There is also a class of real Lorentzian solutions for which the surfaces of homogeneity are timelike and the direction $\dd t$ is spacelike, corresponding to two of the $P_i$ being imaginary and the third being real. The other two cases in which either one or all three of the $P_i$ are taken to be imaginary result in an Urbantke metric of the form $\ii$ times a real Lorentzian metric. These options are consistent with the general solution of the reality conditions given below (\ref{urbantkeformula}).

To deal with these reality conditions in a systematic way, it is useful to rewrite the complex Euler--Lagrange equations (\ref{eqmo1}), (\ref{eqmo2}) and (\ref{eqmo3}) in terms of a vector field $X$ over a real manifold, so that the action of $X$ on functions $f$ of real coordinates on the manifold prescribes their time evolution via $\dot f = X(f)$.
We can use this evolution to test the consistency of the reality conditions and generate further conditions in the inconsistent cases, with some resemblance to Dirac's process for constrained Hamiltonian systems \cite{diracbook}.
We can treat this process as seeking initial conditions on the variables which are preserved under the time evolution given by the Euler--Lagrange equations.

To begin, we write the complex Lagrangian (\ref{Bianchi IX Action Hamiltonian Form}) in terms of real variables by decomposing $C^i = \gamma^i + \ii \Gamma^i$, $P_i = p_i - \ii q_i$ and $\rho = \lambda - \ii \mu$.
Then Lagrangian becomes $L=L_\text{Re} + \ii L_\text{Im}$, with
\begin{subequations}
    \begin{align}
        \label{Real Part Lagrangian}
        & L_\text{Re} = p_i \, \dot \gamma^i + q_i \, \dot \Gamma^i - \lambda H_\text{Re} - \mu H_\text{Im}\,,
        \\
        & L_\text{Im} = -q_i \, \dot \gamma^i + p_i \, \dot \Gamma^i - \lambda H_\text{Im} + \mu H_\text{Re}\,,
    \end{align}
\end{subequations}
where the two constraint functions are given by
\be
H_\text{Re} = \Lambda _\text{P} - \sum_{i=1}^3 \frac{ p_i S^i - q_i T^i }{ (p_i)^2 + (q_i)^2 }\,,\quad
H_\text{Im} = -\sum_{i=1}^3 \frac{ p_i T^i + q_i S^i }{ (p_i)^2 + (q_i)^2 }
\ee
with $S^i$ and $T^i$ defined as cyclic index permutations on 
\be
 S^1 = -\kappa \, \Gamma^1 - \Gamma^2 \Gamma^3 + \gamma^2 \gamma^3        \quad \text{etc.}\,,\quad
 T^1 = \kappa \, \gamma^1 + \gamma^2 \Gamma^3 + \gamma^3 \Gamma^2         \quad \text{etc.}
\ee

Hence we end up with two different real Lagrangians for the same set of real variables.
The real part $L_\text{Re}$ generates the ``real branch'' of the theory and the imaginary part $L_\text{Im}$ the ``imaginary branch''.
The result in appendix \ref{Appendix complex lagrangian} states that since the general complex Lagrangian is holomorphic, the Euler-Lagrange equations of the real and imaginary branches are equivalent; both theories produce the same dynamical evolution on the same variables.
We can then work can work exclusively in the real branch of the theory without loss of generality.
The Euler-Lagrange equations of the real branch are first order and can be computed via
\begin{equation}
   \label{Real branch EL eqn}
    \begin{split}
        \dot \gamma^i = \lambda \frac{ \partial H_\text{Re} }{ \partial p_i } + \mu \frac{ \partial H_\text{Im} }{ \partial p_i }
        \quad & , \quad
        \dot p _i = - \lambda \frac{ \partial H_\text{Re} }{ \partial \gamma^i } - \mu \frac{ \partial H_\text{Im} }{ \partial \gamma^i }
        \\[0.75em]
        \dot \Gamma^i = \lambda \frac{ \partial H_\text{Re} }{ \partial q_i } + \mu \frac{ \partial H_\text{Im} }{ \partial q_i }
        \quad & , \quad
        \dot q _i = - \lambda \frac{ \partial H_\text{Re} }{ \partial \Gamma^i } - \mu \frac{ \partial H_\text{Im} }{ \partial \Gamma^i }\,.
    \end{split}
\end{equation}

The time evolution of arbitrary phase-space functions can be computed as the action of an {\it evolution vector field} $X$ such that $\dot f = X (f)$ with
\begin{equation}
    X = \dot \gamma^i \, \frac{ \partial  }{ \partial \gamma^i } +
    \dot \Gamma^i \, \frac{ \partial  }{ \partial \Gamma^i } +
    \dot p_i \, \frac{ \partial  }{ \partial p_i } +
    \dot q_i \, \frac{ \partial  }{ \partial q_i }\,.
\end{equation}
In addition to the Euler--Lagrange equations (\ref{Real branch EL eqn}) both branches have constraints $H_\text{Re} = 0$ and $H_\text{Im} = 0$, which together define a constraint surface in the configuration space which is the intersection of their zero loci.
This surface is consistent with the time evolution of $X$, that is $\left. X ( H_\text{Re} ) \right \vert_\mathcal{S} = 0$ and $ \left. X ( H_\text{Im} ) \right \vert_\mathcal{S} = 0$.

We can now use the Euler--Lagrange equations derived from $L_\text{Re}$ to apply the reality conditions required for Lorentzian solutions.
First recall that $\rho$ needs to be real, so that $\mu = 0$.
Then, we obtain the constraints $q_i = 0$ {\bf or} $p_i = 0$ respectively for each $i = 1,2,3$.
As we already mentioned this gives four distinct cases: the two homogeneous cases $q_i = 0$ (all real) and $p_i = 0$ (all imaginary) for all $i$, and the mixed cases $\left( q_1 , q_2 , p_3 \right) = 0$ etc.~or $\left( q_1 , p_2 , p_3 \right) = 0$ etc.
Initially none of these cases are self-consistent under evolution by the Euler--Lagrange equations.
For instance, in the all real case with $q_i = 0$ we find
\begin{equation}
    \dot q_1 = -\lambda \left( \frac{\kappa}{ p_1 } + \frac{ \Gamma^3 }{ p_2 } + \frac{ \Gamma^2 }{ p_3 } \right)    \quad \text{etc.}
\end{equation}
which, given that we need $\dot{q}_i=0$, leads to three secondary conditions of the form
\begin{equation}
    \label{Lagrangian secondary constraints}
    \Gamma^i + \Pi^i (p) = 0\,, \qquad
    \Pi^1 (p) := \frac{ \kappa p_2 p_3 }{ 2 } \left( - \frac{ 1 }{ (p_1)^2 } + \frac{ 1 }{ (p_2)^2 } + \frac{ 1 }{ (p_3)^2 } \right)\quad \text{etc.}
\end{equation}
The primary conditions $q_i = 0$ and their secondary conditions $\Gamma^i + \Pi^i = 0$ are (together) consistent; $\dot q_i \approx 0$ and $\left( \Gamma^i + \Pi^i \right)^{ \boldsymbol \cdot } \approx 0$ where $\approx$ indicates equality after the primary and secondary conditions have been applied. Hence, an initial configuration satisfying $q_i = 0$ and $\Gamma^i + \Pi^i = 0$ will continue to satisfy these conditions under evolution by the Euler--Lagrange equations.

This process is similar for the other three cases.
For instance, we can consider the case $p_1 = 0$, $p_2 = 0$, $q_3 = 0$ of one real and two imaginary $P_i$. Here we get secondary conditions
\begin{subequations}
    \label{secondcond1real2im}
    \begin{align}
        & \gamma^1 + \frac{ \kappa \, q_2 p_3 }{2} \left(  \frac{ 1 }{ (q_1)^2 } -\frac{ 1 }{ (q_2)^2 } + \frac{ 1 }{ (p_3)^2 } \right) = 0\,,
        \\[0.5em]
        & \gamma^2 + \frac{ \kappa \, p_3 q_1 }{ 2 } \left( -\frac{ 1 }{ (q_1)^2 } + \frac{ 1 }{ (q_2)^2 }  + \frac{ 1 }{ (p_3)^2 } \right) = 0\,,
        \\[0.5em]
        & \Gamma^3 + \frac{ \kappa q_1 q_2 }{2} \left( \frac{ 1 }{ (q_1)^2 } + \frac{ 1 }{ (q_2)^2 } + \frac{ 1 }{ (p_3)^2 } \right) = 0\,,
    \end{align}
\end{subequations}
which have a very similar form to the conditions in the all real case. 

In all branches of solutions to the reality conditions, some components of the connection $C^i$ can hence be expressed in terms of other dynamical variables. As one might expect, these constraints represent the statement that part of the spin connection is given by the (torsion-free) Levi-Civita connection on homogeneous hypersurfaces, here expressed in the variables of the Pleba{\'n}ski formulation. In the most relevant case where all $P^i$ are real, these hypersurfaces are spacelike, but one can show this also for timelike homogeneous slices in the case of one real, two imaginary $P^i$.

We then need to impose the constraints $ H_\text{Re}=0$ and $ H_\text{Im}=0$ on solutions to the reality conditions and secondary conditions. Here there is a crucial difference between the ``real'' solution branches in which all or exactly one of the $P_i$ are real, and the other ``imaginary'' branches: for the real solutions, $H_\text{Im}=0$ automatically holds once the reality conditions, secondary conditions and $H_\text{Re}=0$ are imposed, and this is not an independent constraint. On the other hand, in the imaginary branches $H_\text{Re}$ and $H_\text{Im}$ cannot both vanish unless $\Lambda=0$, or more generally, unless $\Lambda$ takes an imaginary value as well. As we consider $\Lambda$ to be a fixed, non-zero real fundamental parameter, the imaginary branches hence have no consistent solutions.

\subsection{Hamiltonian dynamics of Lorentzian cosmological solutions}

Whatever branch of the reality conditions one chooses to work in, once a branch is fixed one can view the resulting conditions as constraints in the spirit of Dirac's Hamiltonian formalism. One can then obtain a fully consistent Hamiltonian system, as we will show next.
We should stress that this approach requires fixing one of the four cases we found in the previous subsection  from the start, so the generality of the original complex Lagrangian formulation is lost. The four cases would all lead to different (real) Hamiltonian dynamics.

In the ``all-real'' ($q_i= 0$) and ``one-real, two-imaginary'' ($p_1 = p_2 = q_3 = 0$ etc.) branches, the Hamiltonian formalism is obtained from the real part of the Lagrangian, $L_\text{Re}$, whereas the other ``all-imaginary'' and ``two-real, one-imaginary'' cases are based on the imaginary part $L_\text{Im}$. 
This is related to the fact that in the first two branches the Urbantke metric is real while in the other two it is imaginary. Consistent dynamics for real, non-zero $\Lambda$ can only be obtained in the all-real and one-real, two-imaginary cases, so we will not discuss the other cases further.

In the most relevant all-real case, we initially have a phase space $\mathcal{P} = \R^{12}$ with coordinates $\left( \gamma^i , \Gamma^i , p_i , q_i \right)$ for $i = 1,2,3$, and symplectic form $\Omega = \dee p_i \wedge \dee\gamma^i + \dee q_i \wedge \dee \Gamma^i$.
This form generates a Poisson bracket $\{\cdot,\cdot\}$ on the algebra $C^\infty \left( \mathcal{P} \right)$ satisfying $\left\{ \gamma^i , p_j \right\} = \left\{ \Gamma^i , q_j \right\} =  \delta^i _j$.
Our initial system has a pair of constraints $H_\text{Re} = 0$ and $H_\text{Im} = 0$ which satisfy $\left\{ H_\text{Re} , H_\text{Im} \right\} \approx 0$ making them first class.
The Euler--Lagrange equations (\ref{Real branch EL eqn}) can be written using the bracket as
\begin{equation}
    \label{naive time evolution}
    \dot f \approx \lambda \left\{ f , H_\text{Re} \right\} + \mu \left\{ f , H_\text{Im} \right\}
\end{equation}
where $\lambda$ and $\mu$ are arbitrary phase space functions.

We would now like to impose reality conditions. First, in agreement with what we did in the Lagrangian setting, we will assume that the trace reality condition $\text{Re} \left( \Sigma^i \wedge \Sigma_i \right) = 0$ requires $\mu=0$ in our evolution equations.
This is a condition on a Lagrange multiplier rather than on phase space functions, and means that $H_\text{Im}$ does not contribute to the time evolution.
Next we add the wedge reality conditions $F^i \wedge \overline{ F^j } = 0$ to our theory.
In our solution branch we assume that these are imposed as $q_i \approx 0$. 
We can add these constraints to the Hamiltonian in the usual way, so that
\begin{equation}
    \dot f \approx \lambda \left\{ f , H_\text{Re} \right\} + u^i \left\{ f , q_i \right\}
\end{equation}
where $u^i$ are arbitrary functions of the variables.
We now examine the consistency conditions $\dot q_i \approx 0$ which are cyclic index permutations on
\begin{equation}
    -\lambda \left( \frac{ \kappa }{ p_1 } + \frac{ \Gamma^3 }{ p_2 } + \frac{ \Gamma^2 }{ p_3 } \right) \approx 0
    \quad \text{etc.}
\end{equation}
These conditions yield secondary constraints 
\begin{equation}
    \frac{ \kappa }{ p_1 } + \frac{ \Gamma^3 }{ p_2 } + \frac{ \Gamma^2 }{ p_3 } = 0
    \quad \text{etc.}
    \quad \Rightarrow \quad
    \Gamma^1 + \frac{ \kappa \, p_2 p_3 }{ 2 } \left( - \frac{ 1 }{ (p_1)^2 } + \frac{ 1 }{ (p_2)^2 } + \frac{ 1 }{ (p_3)^2 } \right) = 0
    \quad \text{etc.}\,,
\end{equation}
i.e., exactly the constraints (\ref{Lagrangian secondary constraints}) found using Lagrangian dynamics.
We add these secondary constraints to our theory which brings us up to a total of eight constraints.
However, the six mutually independent constraints $q_i = 0$ and $\Gamma^i + \Pi^i = 0$ together imply the constraint $H_\text{Im} = 0$, so we can drop the constraint $H_\text{Im}$ from the theory without harm leaving us with only seven constraints.
Going into the next round of consistency checks, the evolution equations are
\begin{equation}
    \dot f \approx \lambda \left\{ f , H_\text{Re} \right\} + u^i \left\{ f , q_i \right\} + v_i \left\{ f , \Gamma^i + \Pi ^i(p) \right\}
\end{equation}
where $v_i$ are arbitrary functions of the variables.
We now check the consistency of our seven constraints under this time evolution.
We find $\dot H _\text{Re} \approx 0$ automatically, and the constraint is self-consistent.
However, we have $\dot q_i \approx -v_i$ and $\left( \Gamma^i + \Pi^i(p) \right) ^{ \boldsymbol \cdot } \approx -u^i$, so consistency requires to fix the Lagrange multipliers $u^i \approx 0$ and $v_i \approx 0$. These constraints are second class. There are no further constraints in the theory.

At this stage, our evolution equations $\dot f \approx \lambda\left\{ f , H_\text{Re} \right\}$ could be extended to $\dot f \approx \left\{ f , H_\text{T} \right\}$ with $H_\text{T} = \lambda H_\text{Re} + w^A \tau_A {}^a \phi_a$, where $w^A$ are Lagrange multipliers and $A$ indexes over the independent solutions of the homogeneous equations $\tau_A {}^a \left\{ \phi_a , \phi_b \right\} \approx 0$, and where $\phi_a$ are the seven constraints in the theory. However, the only independent solution satisfies $\tau^a \phi_a \approx H_\text{Re}$, so we already have the most general consistent time evolution on our constrained system.

We can now continue with the standard Dirac algorithm \cite{diracbook}. 
Denoting the six second class constraints $\left( q_i, \Gamma^i + \Pi ^i(p) \right)$ by the collective label $\chi_\alpha$ with $\alpha = 1 , \ldots , 6$, the matrix of brackets $W_{ \alpha \beta } = \left\{ \chi_\alpha , \chi_\beta \right\}$ is non-degenerate on the constraint surface, so we have already identified the maximal number (one) of first class constraints.
Using the inverse components $W^{ \alpha \beta }$ satisfying $W^{ \alpha \gamma } W_{ \gamma \beta } = \delta^\alpha _\beta$, we can construct a Dirac bracket which will take the place of the Poisson bracket:
\begin{equation}
    \left\{ f , g \right\}^* = \left\{ f , g \right\} - \left\{ f , \chi_\alpha \right\} W^{ \alpha \beta } \left\{ \chi_\beta , g \right\}\,.
\end{equation}
All of the constraints in the theory are first class with respect to this bracket:
the Dirac bracket satisfies $\left\{ f , \chi_\alpha \right\}^* \approx 0$ for all phase space functions $f$ and second class constraints $\chi_\alpha$, and $\left\{ f , \mathcal{C} \right\}^* \approx \left\{ f , \mathcal{C} \right\}$ for all phase space functions $f$ and a first class constraint $\mathcal{C}$.
Defining an extended Hamiltonian $H_\text{E} = \lambda H_\text{Re} + U^\alpha \chi_\alpha$, we can write down evolution equations in terms of the Dirac bracket as
\begin{equation}
    \dot f = \left\{ f , H_\text{E} \right\}^* \approx \lambda \left\{ f , H_\text{Re} \right\}^* + U^\alpha \left\{ f , \chi_\alpha \right\}^*
\end{equation}
Since $\left\{ f , H_\text{E} \right\}^* \approx \lambda \left\{ f , H_\text{Re} \right\}$, this definition of time evolution is equivalent to our previous definition.
We can use the constraints $\chi_\alpha$ to rewrite the first class constraint $H_\text{Re}$ in terms of only the variables $\gamma^i$ and $p_i$.
This rewritten constraint $\mathcal{H}$ can be written as $\mathcal{H} = \mathcal{H}_{ ( - ) } + \kappa^2 \, \mathcal{H}_\text{BG}$, where
\be
    \label{Real Hamiltonian constraint}
         \mathcal{H}_{ ( \pm ) } = \Lambda _\text{P} \pm \left( \frac{ \gamma^2 \gamma^3 }{ p_1 } + \frac{ \gamma^3 \gamma^1 }{ p_2 } + \frac{ \gamma^1 \gamma^2 }{ p_3 } \right)\,,\quad
         \mathcal{H}_\text{BG} = \frac{1}{2} \sum_{i=1}^3 \left( \frac{ p_1 p_2 p_3 }{ 2 (p_i)^4 }  - \frac{ (p_i)^2 }{ p_1 p_2 p_3 } \right)\,.
\ee
This decomposition illustrates the effect of background curvature (given by the parameter $\kappa$) on the dynamical evolution of the spatial geometry. One can check that (\ref{Real Hamiltonian constraint}) represents the Hamiltonian constraint of the Lorentzian Bianchi IX model in general relativity with cosmological constant $\pm\Lambda$ (with the sign depending on the sign of the product $p_1 p_2 p_3$), for example by starting from Ashtekar variables (see Appendix \ref{bianchiIXapp}) or from the metric representation.

Explicitly, if we denote the phase space coordinates collectively by $X^A=\left( \gamma^i, \Gamma^i, p_i, q_i \right)$,
the Dirac bracket can be given in matrix form as
\begin{equation}
    \label{DB lookup table}
    \{X^A,X^B\}^*=\mat{
    0 & {\bf A} & {\boldsymbol 1}  & 0 \\
    - {\bf A}^\text{T} & 0 & 0 & 0 \\
    - {\boldsymbol 1} & 0 & 0 & 0 \\
    0 & 0 & 0 & 0
    }
\end{equation}
where each entry is a $3 \times 3$ matrix where $\boldsymbol 1$ denotes the identity while 0 denotes the empty matrix.
The symmetric matrix ${\bf A}$ has components
\begin{equation}
    A^{ij} = - \frac{ \partial \Pi ^j }{ \partial p_i }\,.
\end{equation}
It is easily seen by direct calculation that $A^{ij}$ is symmetric under the exchange of $i$ and $j$.

From \cite{henneaux2020quantization} we have a further useful property for the Dirac bracket:
\begin{equation}
    \label{Henneaux Dirac bracket result}
    \left\{ F , G \right\}^* \vert_{ \chi_\alpha = 0 } = \left\{ F \vert_{ \chi_\alpha = 0 } , G \vert_{ \chi_\alpha = 0 } \right\}^*\,,
\end{equation}
where $ \vert_{ \chi_\alpha = 0 }$ denotes restriction to the second class constraint surface where $\chi_\alpha = 0$.
We can define this restriction by $F \vert_{ \chi_\alpha = 0 } ( \gamma^i , p_i ) := F \left( \gamma^i , - \Pi^i (p) , p_i , 0 \right)$ noting that $( \gamma^i , p_i )$ can be used as coordinates on the second class constraint surface $\mathcal{P}_\text{R}=\R^6$.
The lookup table (\ref{DB lookup table}) reveals that $\{ \gamma^i , p_j \}^* = \delta^i _j$ is just the standard Poisson bracket $\{ \cdot , \cdot \}_0$ on $\mathcal{P}_\text{R}$. Then from (\ref{Henneaux Dirac bracket result}) the following diagram commutes:
\begin{equation}
    \begin{array}{ccc}
        C^\infty ( \mathcal{P} ) \times C^\infty ( \mathcal{P} ) & \overset{ \left\{ \cdot , \cdot \right\} ^* }{ \xrightarrow{ \hspace*{2cm} } } & C^\infty ( \mathcal{P} )
        \\
        \bigg \downarrow J & & \bigg \downarrow J
        \\
        C^\infty ( \mathcal{P}_\text{R} ) \times C^\infty ( \mathcal{P}_\text{R} ) & \overset{ \left\{ \cdot , \cdot \right\}_0 }{ \xrightarrow{ \hspace*{2cm} } } & C^\infty ( \mathcal{P}_\text{R} )
    \end{array}
\end{equation}
Here $J : F \mapsto F \vert_{ \chi_\alpha }$ sends a phase space function to its restricted version.
For peace of mind, we can confirm by direct calculation that
\begin{equation}
    \begin{split}
        \{ F \vert_{ \chi_\alpha } , G \vert_{ \chi_\alpha } \}_0 & \overset{!}{=} \frac{ \partial F }{ \partial \gamma^i } \left( - \frac{ \partial G }{ \partial \Gamma^j } \frac{ \partial \Pi^j }{ \partial p_i } + \frac{ \partial G }{ \partial p_i } \right) - \frac{ \partial G }{ \partial \gamma^i } \left( - \frac{ \partial F }{ \partial \Gamma^j } \frac{ \partial \Pi^j }{ \partial p_i } + \frac{ \partial F }{ \partial p_i } \right)
        \quad \bigg \vert_{ \chi_\alpha = 0 }
        \\[0.75em]
        & = \frac{ \partial F }{ \partial \gamma^i } \frac{ \partial G }{ \partial p_i } - \frac{ \partial G }{ \partial \gamma^i } \frac{ \partial F }{ \partial p_i }
        -
        \frac{ \partial \Pi ^j }{ \partial p_i } \left( \frac{ \partial F }{ \partial \gamma^i } \frac{ \partial G }{ \partial \Gamma^j } - \frac{ \partial G }{ \partial \gamma^i } \frac{ \partial F }{ \partial \Gamma^j } \right)
        \quad \bigg \vert_{ \chi_\alpha = 0 }
        \\
        & = \left\{ F , G \right\}^* \vert_{ \chi_\alpha = 0 }\,.
    \end{split}
\end{equation}
We can hence define a Hamiltonian theory over a lower-dimensional phase space $\mathcal{P}_\text{R}$ that contains the same dynamical content as the theory on $\mathcal{P}$.
To do this, we promote the second class constraints to strong conditions and drop the variables $\Gamma^i , q_i$ from the theory. $\mathcal{P}_\text{R}$ has the standard Poisson bracket, and the resulting theory is the standard Hamiltonian formulation of the Lorentzian Bianchi IX model.

We finally note that this reduced Hamiltonian theory on $\mathcal{P}_\text{R}$ can be generated from the action
\begin{equation}
    \label{Reduced Lorentzian action}
    S_\text{R} \left[ \gamma , p , \lambda \right] = \int \dee t\left(p_i \, \dot \gamma^i - \lambda \left( \mathcal{H}_{ ( - ) } + \kappa ^2 \, \mathcal{H}_\text{BG} \right)\right)
\end{equation}
which can be directly obtained from the first order action (\ref{Chiral first order action}) with the substitutions
\begin{subequations}
    \label{field definitions Bianchi IX}
    \begin{align}
        & A^i = V^{-1/3} \left( \ii \gamma^i + \Pi ^i (p) \right) e^i\,,
        \\[0.75em]
        & M^{11} = \frac{ \kappa \left( \ii \gamma^1 + \Pi^1(p) \right) - \left( \ii \gamma^2 + \Pi^2(p) \right) \left( \ii \gamma^3 + \Pi^3(p) \right) }{ \ell_\text{P}^2\, p_1 }
        \quad \text{etc.}\,,
        \\[0.75em]
        & \nu = 2\ii V^{-1} \ell_\text{P}^4\, \lambda \, \varepsilon_e \wedge \dee t\,,
    \end{align}
\end{subequations}
setting the off-diagonal elements of $M^{ij}$ to zero.

A similar analysis can be repeated for branch characterised by one real and two imaginary $P_i$. For instance, assuming primary constraints $p_1 = 0$, $p_2 = 0$, $q_3 = 0$ will lead to the secondary constraints given in (\ref{secondcond1real2im}). Furthermore, for this case the Hamiltonian constraint is
\be
    \label{Real Hamiltonian constraint 2real 1imaginary}
         \mathcal{H} = \Lambda _\text{P} + \left( \frac{ \Gamma^2 \gamma^3 }{ q_1 } + \frac{ \gamma^3 \Gamma^1 }{ q_2 } + \frac{ \Gamma^1 \Gamma^2 }{ p_3 } \right)+ \frac{\kappa^2}{2} \left(- \frac{ q_1 q_2 }{ 2 (p_3)^3 }- \frac{ q_2 p_3}{ 2 (q_1)^3 }- \frac{ q_1 p_3 }{ 2 (q_2)^3 }  - \frac{ q_1 }{ q_2 p_3 }- \frac{ q_2 }{ q_1 p_3 } + \frac{ p_3 }{ q_1 q_2} \right)\,.
\ee
This case then also leads to a well-defined Hamiltonian system, which can be studied in its own right. The corresponding Urbantke metric (\ref{Urbantke metric}) gives timelike surfaces of homogeneity and a spacelike direction ${\rm d}t$ of nontrivial evolution, so this is not a usual cosmological Bianchi model. From the perspective of the Euclidean theory we will study below, it can be seen as a ``Wick rotation'' in a spatial, rather than timelike direction.

\section{Diagonal Bianchi I model}
\label{Bianchi I section}

Our analysis of the Bianchi IX model in the first order connection theory was quite involved; the main issues were identifying Lorentzian solutions in the generally complex theory and studying the dynamics of the different solution branches in Hamiltonian form. We saw that if one wants to include the reality conditions as constraints in a Hamiltonian description, they and their secondary consistency conditions are second class; one can pass to a reduced phase space description with only a single (Hamiltonian) constraint once a solution branch of the reality conditions is identified. In the application to cosmology, there is only one interesting branch, in which in the Urbantke metric the homogeneous hypersurfaces are spacelike and the direction of evolution is timelike.

A somewhat simpler anisotropic model is obtained by assuming that the homogeneous submanifolds ${\bf S}$ in the decomposition $\mathcal{M} = \R \times {\bf S}$ are diffeomorphic to flat $\R^3$; the group of isometries acting on each of these leaves is then an Abelian group of translations. This is the Bianchi I model, which was studied for a generalised class of pure connection theories of gravity in \cite{KirillBianchiI}.

The Cartan frame $\underline{e}^{\bf a}$ now satisfies $\dee \underline{e}^{\bf a} = 0$; a simple example would be $\underline{e}^{\bf a} = \dee x^i$ in Cartesian coordinates $x^i$ on $\R^3$. Again, we have a volume form $\varepsilon_{\underline{e}} := \underline{e}^1 \wedge \underline{e}^2 \wedge \underline{e}^3$ and, in contrast to the case of $S^3$, the total volume of space is now infinite. Hence, restriction to a ``fiducial cell'' or more generally a compact and connected submanifold $\mathcal{V}_0 \subseteq \R^3$ is necessary. As before we define $V_0 := \int_{\mathcal{V}_0} \varepsilon_{\underline{e}}$. We are still assuming a diagonal connection $A^i = C^i (t) \, \delta^i _{\bf a} \, \underline{e}^{\bf a}$ (no sum over $i$), so that
\begin{equation}
    A^i = \ii V_0^{-1/3} C^i \, \underline{e}^i\,,
    \qquad
    F^1 = \ii V_0^{-1/3} \dot C^1 \, \dee t \wedge \underline{e}^1 - V_0 ^{-2/3} C^2 C^3 \, \underline{e}^2 \wedge \underline{e}^3
    \quad \text{etc.}
\end{equation}

Again we can take the auxiliary matrix to be diagonal, with entries
\begin{equation}
    M^{11} = \frac{  C^2 C^3 }{\ell_\text{P} ^{2} P_1 }
    \quad \text{etc.}\,,
\end{equation}
and parametrise the auxiliary form as $\nu = 2 \ii\, V_0 ^{-1} \ell_\text{P}^4\, \rho \, \varepsilon_{\underline{e}} \wedge \dee t$ where $\rho$ is a complex scalar density, transforming under time reparametrisations according to $\rho(t)\,\dee t=\rho'(t')\,\dee t'$. Inserting all these fields into the first order action (\ref{Chiral first order action}) leads to a reduced action
\begin{equation}
   \label{Bianchi I action}
    S \left[ C , P , \rho \right] = \int \dee t\left[P_i \, \dot C^i - \rho \left( \Lambda_\text{P} -  \frac{ C^2 C^3 }{ P_1 } -  \frac{ C^1 C^3 }{ P_2 } -  \frac{ C^1 C^2 }{ P_3 }  \right) \right]\,.
\end{equation}
It is now clear that this reduced theory is equivalent to the reduced theory describing the Bianchi IX model when we take $\kappa\rightarrow 0$. None of the equations described in section \ref{Diagonal symmetric models section} become singular or ill-defined in this limit, so that we can take over all results obtained there and set $\kappa=0$. The global topological differences between $S^3$ and $\R^3$ play no role at the level of (\ref{Bianchi I action}).

The reality conditions still imply that $\text{Im} \left( \rho \right)=0$ and $\text{Im} \left( P_i{}^2 \right)=0$ for all $i$, with the latter condition defining various solution branches depending on whether one takes the solution to be $q_i=0$ (i.e., $P_i$ is real) or $p_i=0$ (i.e., $P_i$ is imaginary). The secondary conditions derived in section \ref{Lorentzian Bianchi}, however, simplify greatly. For instance, for the all-real case we found $\Gamma^i+\Pi^i(p)=0$ in (\ref{Lagrangian secondary constraints}); as $\kappa\rightarrow 0$, $\Pi^i(p)\rightarrow 0$ and these conditions simply reduce to $\Gamma^i=0$, with $\Gamma^i$ the conjugate variable to $q_i$. This is as we would
expect, as $\Gamma^i$ represents the Levi-Civita connection on spatial hypersurfaces which are flat. The reality conditions and secondary constraints hence become the simplest possible form of second-class constraints, requiring both variables of a conjugate pair to vanish. These constraints can be ``solved'' by simply removing these variables from the theory. 

The Bianchi I model hence admits a particularly simple reduction to a real Lorentzian theory: all we need to do is to demand that all dynamical variables appearing in (\ref{Bianchi I action}) are real-valued. One might think that this is an off-shell solution of the reality conditions, not requiring knowledge of dynamical equations. However, this is not really the case: the original form of the reality conditions of wedge type is $\text{Im} \left( \overline{ C^2 C^3 } \, \dot C^1 \right) = 0$ and permutations, and it is only when we insert the dynamical equations for $\dot{C}^i$ that these reduce to a simple algebraic form for $C^i$ and $P_i$. This is the point we already made below (\ref{realitycond}) in the more general setting.

The Bianchi I model is also simple enough to be solved analytically. The equations of motion in this case are
\begin{equation}
    \dot C^1 = \rho \, \frac{C^2 C^3 }{ (P_1)^2 }
        \quad \text{etc.}  \,,\qquad
    \dot P_1 = \rho \left( \frac{ C^3 }{ P_2 } + \frac{ C^2 }{ P_3 } \right)
        \quad \text{etc.}
\end{equation}
We can solve the last three equations algebraically for $C^i$ to find
\be
   \label{connection solution}
     C^1 = \frac{P_2P_3}{2\rho} \left(-\frac{\dot{P}_1}{P_1} +\frac{\dot{P}_2}{P_2} +\frac{\dot{P}_3}{P_3}\right) \quad \text{etc.}
\ee
The Hamiltonian constraint then becomes
\be
  \Lambda_\text{P}+\frac{P_1P_2 P_3}{4\rho^2}\left(\frac{(\dot{P}_1)^2}{(P_1)^2}-2\frac{\dot{P}_2\dot{P}_3}{P_2 P_3}+\text{permutations}\right) = 0\,.
\ee
We may rewrite this in the gauge $\rho^2= P_1P_2P_3$ and writing $P_i(t)=\exp(\tau_i(t))$ as
\be
\label{bianchiIcons}
  \Lambda_\text{P}+\frac{1}{4}\left(\dot{\tau}_1^2+\dot{\tau}_2^2+\dot{\tau}_3^2-2\dot\tau_1\dot\tau_2-2\dot\tau_1\dot\tau_3-2\dot\tau_2\dot\tau_3\right)=0\,.
\ee
The dynamical equations for $C^i$ become, using (\ref{connection solution}),
\be
\label{connectionIeq}
 -\ddot\tau_1+\ddot\tau_2+\ddot\tau_3-\dot\tau_1\left(\dot\tau_2+\dot\tau_3\right)+\dot\tau_2^2 + \dot\tau_3^2 = 0 \quad\text{etc.}\,,
\ee
that is, three first-order differential equations for the quantities $\dot\tau_i$. The sum of these three equations can be written as
\be
\ddot\tau_1 + \ddot\tau_2 + \ddot\tau_3 + \frac{1}{2}\left(\dot\tau_1+\dot\tau_2+\dot\tau_3\right)^2 = 6\Lambda_\text{P}
\ee
using the constraint (\ref{bianchiIcons}). This is a Riccati equation for the quantity $\sum_i\dot\tau_i$, with solution
\be
\dot\tau_1+\dot\tau_2+\dot\tau_3 = 2\sqrt{3\Lambda_\text{P}}\coth\left(\sqrt{3\Lambda_\text{P}}(t-t_0)\right)\,.
\ee
Furthermore, by adding two of the equations (\ref{connectionIeq}) and using the constraint again, we find
\be
\ddot{\tau}_1+\frac{1}{2}\dot{\tau}_1(\dot\tau_1+\dot\tau_2+\dot\tau_3)=2\Lambda_\text{P}\quad\text{etc.}
\ee
which can now be solved to get
\be
\label{taudotsolutions}
\dot{\tau}_i = b_i\, {\rm cosech}\left(\sqrt{3 \Lambda_\text{P}}(t-t_0)\right) +2\sqrt{\frac{\Lambda_\text{P}}{3}}\coth\left(\sqrt{3\Lambda_\text{P}}(t-t_0)\right)
\ee
where the $b_i$ are integration constants satisfying $b_1+b_2+b_3=0$. Substituting these solutions into (\ref{bianchiIcons}) then gives a second constraint,
\be
b_1b_2+b_1b_3+b_2b_3=-\Lambda_{\text{P}}\,.
\ee
Integrating and exponentiating the solutions  (\ref{taudotsolutions}) finally yields
\be
P_i = \tilde{P}_i \tanh^{r_i}\left(\frac{\sqrt{3\Lambda_\text{P}}}{2}(t-t_0)\right)\sinh^{2/3}\left(\sqrt{3\Lambda_\text{P}}(t-t_0)\right)
\ee
with the rescaled exponents $r_i$ now satisfying $\sum_i r_i=0$ and $r_1r_2+r_1r_3+r_2r_3=-\frac{1}{3}$. From this expression one may reconstruct the Urbantke metric via (\ref{Urbantke metric}) and check that these solutions reproduce the Bianchi I solutions with cosmological constant originally given by Kasner \cite{Kasner} (see also \cite{exactsolutions}).

\section{Relation to Euclidean theory}
\label{euclideansec}

The fundamental reason why the chiral Pleba\'{n}ski formulation of gravity and its descendants employ complex fields, leading to the need for reality conditions to obtain Lorentzian solutions, is that the Hodge star operator in Lorentzian signature squares to $-1$; hence self-dual and anti-self-dual objects must have eigenvalues $\pm\ii$. In Riemannian signature, the Hodge star squares to $+1$ and the corresponding eigenvalues are $\pm 1$. No complexification is necessary in this case. This fact allows studying a Euclidean definition\footnote{Here we are following the usual physics terminology of ``Euclidean'' definitions of gravity as theories of Riemannian geometries, even though these geometries do not correspond to Euclidean, i.e., flat space.} of the Pleba\'{n}ski and chiral first order formulations of general relativity based on real fields only. We then expect there to be a correspondence between solutions in Euclidean and Lorentzian signature, which can be understood as a ``Wick rotation'' at least in simple situations such as homogeneous cosmology, where there is an unambiguous notion of the time coordinate to be ``rotated.'' Understanding such a correspondence provides an alternative to the approach followed so far, in which we start with entirely complex fields and then select those solutions which correspond to Lorentzian geometries. While the Pleba\'{n}ski formalism seems to be somewhat unique in using real variables in Euclidean signature and complex variables in Lorentzian signature, the general idea of defining a theory initially in Euclidean signature and obtaining Lorentzian answers only by analytic continuation afterwards is a standard procedure (and was followed, e.g., in the first systematic path integral approach for gravity \cite{EuclideanQG}).

For this section, we define the Hodge star in terms of a volume form $\varepsilon = E^1\wedge E^2 \wedge E^3\wedge E^0$, so that
\be
\star ( E^0 \wedge E^1 ) = - E^2 \wedge E^3\,,\quad\star ( E^2 \wedge E^3 ) = - E^0 \wedge E^1\,,
\ee
and so on. $\Sigma^1 := E^0 \wedge E^1 - E^2 \wedge E^3$ etc.~are then a basis of self-dual 2-forms. 
Again, this basis arises from a related duality on real bivectors (antisymmetric $\R^4 \otimes \R^4$ tensors), which form a representation of the algebra $\mathfrak{so}(4)$.
The dual $\ast$ acts in the same way as before, except the Latin indices $I,J$ etc.~are now raised and lowered with the Kronecker 4-deltas $\delta^{IJ} , \delta_{IJ}$.
In this case $\ast$ has eigenvalues $\pm 1$, and self-dual bivectors satisfy $\ast B^{IJ} = B^{IJ}$.
The self-dual projector is defined $P_+ ^{IJ} {}_{KL} = (1/2) \left( \delta^{ [I } _K \delta^{ J] } _L + (1/2) \epsilon^{IJ} {}_{KL} \right)$ and sends bivectors to their self-dual parts $B^{IJ} \to B_+ ^{IJ}$.
We perform a change of basis $B_+ ^{IJ} \to B^i = 2 B_+ ^{0i}$, then $\Sigma^i = 2 ( E \wedge E )_+ ^{0i}$ in parallel with the Lorentzian case.
The self-dual part of the connection is defined as before; we obtain a real $SO(3)$ connection $A^i$ related to the usual spin connection $\omega^I{}_J$ by $A^1 = {\omega^0}_1 - {\omega^2}_3$, etc., see \cite{KrasnovBook}.

The Euclidean analogue of the chiral first order action (\ref{Chiral first order action}) reads
\begin{equation}
    S_\text{EFO} [ A , M , \nu ] = \frac{1}{ 2 \, \ell_\text{P}^2 } \int M^{-1}_{ ij } F^i \wedge F^j + \left( \tr M - \Lambda \right) \nu
\end{equation}
where the key difference is the lack of an overall factor $-\ii$ present in the Lorentzian version. All fields are now required to be real-valued, so $A^i$ are the component 1-forms of a real $SO(3)$ connection, with $F^i$ its corresponding curvature 2-forms. 

We can now study a diagonal Bianchi IX model in analogy to the Lorentzian discussion of section \ref{Diagonal symmetric models section}. The homogeneous and diagonal connection ansatz and expression for the curvature replacing (\ref{Connection and curvature ansatz}) are now
\begin{equation}
    A^i = V ^{-1/3} C^i \, e^i\,,
    \qquad
    F^1 = V ^{-1/3} \dot C^1 \, \dee t \wedge e^1 - V ^{-2/3} \left(\kappa C^1 - C^2 C^3 \right) \, e^2 \wedge e^3
    \quad \text{etc.}\,;
\end{equation}
 parametrising the auxiliary matrix as 
\begin{equation}
    M^{11} = \frac{\kappa C^1 - C^2 C^3 }{\ell_\text{P}^2 P_1 }
    \quad \text{etc.}
\end{equation}
and the auxiliary form as $\nu = 2V^{-1}\ell_\text{P}^4\,\rho\,\epsilon_e\wedge\dee t$ then leads to a reduced action
\begin{equation}
    \label{Euclidean Hamiltonian form}
    S_\text{E} \left[ C , P , \rho \right] = \int \dee t\left[P_i \, \dot C^i - \rho \left( \ell_\text{P}^2 \Lambda + \left( \frac{ C^2 C^3 }{ P_1 } + \text{permutations} \right) - \kappa \sum_{i=1} ^3 \frac{ C^i }{ P_i } \right) \right]\,.
\end{equation}
We repeat that all fields are now real, so there is no need to impose any additional conditions such as reality conditions. The form of (\ref{Euclidean Hamiltonian form}) immediately leads to the definition of a Hamiltonian system on phase space $\mathcal{P}_\text{E}=\R^6$ with coordinates $(C^i,P_i)$ and Poisson bracket $\{C^i,P_j\}=\delta^i_j$. There is a single Hamiltonian constraint
\be
         \mathcal{H}_\text{E} = \Lambda_\text{P} + \left( \frac{ C^2 C^3 }{ P_1 } + \text{permutations} \right) - \kappa \sum_{i=1} ^3 \frac{ C^i }{ P_i } 
\ee
with $\Lambda_\text{P}:=\ell_\text{P}^2 \Lambda$ the value of the cosmological constant in Planck units as defined previously.

In order to clarify the correspondence between this theory and the previously defined reduced Lorentzian action (\ref{Reduced Lorentzian action}) we need to apply a transformation to the connection $C^i$, motivated by its interpretation in the Euclidean Pleba\'{n}ski formulation of gravity. As in section \ref{reconstruct metric}, this interpretation is obtained by considering the Urbantke metric, now defined by (cf.~\ref{urbantkeformula})
\be
\tilde{\varepsilon}_\Sigma = \frac{1}{6} \Sigma_\text{E} ^i \wedge (\Sigma_\text{E})_i\,,\qquad g_E \left( \xi , \eta \right) \, \tilde{\varepsilon}_\Sigma = -\frac{1}{6} \epsilon_{ijk} \, i_\xi \Sigma_\text{E}^i \wedge i_\eta \Sigma_\text{E}^j \wedge \Sigma_\text{E}^k
\ee
with the Euclidean Pleba\'{n}ski 2-forms defined by $\Sigma^i_\text{E}=(M^{-1})^{ij}F_j$ for the real curvature 2-forms $F_i$. Explicitly, in this case we find (cf.~\ref{simplesigmaf})
\begin{equation}
    \Sigma_E ^1 = -V^{-2/3} \ell_\text{P}^2 \left( \frac{V^{1/3} \rho }{ P_1 } \, \dee t \wedge e^1 + P_1 \, e^2 \wedge e^3 \right)\quad \text{etc.}\,,
\end{equation}
resulting in the Euclidean Urbantke metric
\begin{equation}
    g_E = \frac{ \ell_\text{P}^2\, \rho^2 }{ P_1 P_2 P_3 } \, \dee t \otimes \dee t + V ^{-2/3} \ell_\text{P}^2 \, P_1 P_2 P_3 \, \sum_{i=1}^3 \frac{1}{ (P_i)^2 } \, e^i \otimes e^i\,.
\end{equation}
As one might have expected, this differs from our previous result (\ref{Urbantke metric}) by a minus sign in front of the $\dee t \otimes \dee t$ part of the metric, so that we obtain a metric of positive definite signature for $P_1 P_2 P_3 >0$ and negative definite signature for $P_1 P_2 P_3 < 0$.

We find the $SO(3)$ Levi-Civita connection on constant $t$ slices, corresponding to the physical triad $E^i = V ^{-1/3} \ell_\text{P}\, \frac{\sqrt{|P_1 P_2 P_3|}}{P_i} \, e^i$, to be given by $\Gamma_a^i = -V^{-1/3} \Pi^i(P) e^i_a$, with $\Pi^i(P)$ as in (\ref{Lagrangian secondary constraints}). Again, this is an expected result; this metric on spatial hypersurfaces has the same form in Euclidean and Lorentzian signature, so the Levi-Civita connection on these slices must be the same too.

We can then use the fact that the $SO(3)$ connection $A^i$ in the Euclidean Pleba\'{n}ski theory is related to the usual extrinsic curvature and spatial Levi-Civita connection parts of an $SO(4)$ spin connection, which we may denote by $K^i$ and $\Gamma^i$ respectively, as $A^i=K^i-\Gamma^i$. In other words, the extrinsic curvature, which is the dynamical part of the connection for a spatially homogeneous cosmology, can be isolated by defining $K^i:=A^i+\Gamma^i$ with $\Gamma^i$ obtained from the Urbantke metric. In our case, this amounts to a coordinate transformation
\be
\gamma^i = C^i - \Pi^i(p)\,,\qquad p_i = P_i\,.
\ee
This is obviously a canonical transformation, i.e., the new variables $\gamma^i$ and $p_i$ are again canonically conjugate. (This canonical transformation relating extrinsic curvature and an $SO(3)$ connection is also the basis of the Ashtekar--Barbero formulation of Lorentzian general relativity \cite{thiemanbook}.) Written in terms of  $(\gamma^i,p_i)$, the Hamiltonian constraint becomes $\mathcal{H}_\text{E} = \mathcal{H}_{(+)}+\kappa^2\mathcal{H}_\text{BG}$ with $\mathcal{H}_{(+)}$ and $\mathcal{H}_\text{BG}$ defined in (\ref{Real Hamiltonian constraint}). 

We are now in a position to define a kind of Wick rotation (i.e., a transformation involving a replacement of real by purely imaginary variables) between this theory and the different Lorentzian branches of the reality conditions studied for the complex theory. Namely, consider the Hamiltonian equations of motion of the Euclidean theory,
\be
\frac{\dee f}{\dee t}=\rho\{f,\mathcal{H}_\text{E}\}
\ee
where $f$ is an arbitrary function on the Euclidean phase space $\mathcal{P}_{\text{E}}$. Then notice that a complex transformation $\gamma^i\mapsto \tilde\gamma^i=\ii\gamma^i$ maps $\mathcal{H}_\text{E}$ to the Lorentzian constraint $\mathcal{H}=\mathcal{H}_{(-)}+\kappa^2\mathcal{H}_\text{BG}$ while changing the symplectic structure by an overall constant factor, $\{\cdot,\cdot\}\mapsto \ii\{\cdot,\cdot\}$. Absorbing this constant into the Lagrange multiplier, we may conclude that if $(\gamma^i(t),p_i(t),\rho(t))$ defines a solution to the Euclidean theory then $(\ii\gamma^i(t),p_i(t),\ii\rho(t))$ formally defines a solution to the Lorentzian theory. The map is obviously invertible, i.e., any Lorentzian solution likewise maps into a Euclidean one. 

This argument generalises to any pair of theories each characterised by a single Hamiltonian constraint $H_i$, where one can find a (complex) transformation on the variables $x^\mu \mapsto x' {}^\mu (x)$ satisfying $H_1 (x) = H_2 \left( x' (x) \right)$ and $\left. \{ x' {}^\mu , x' {}^\nu \} \right \vert _{ x } = \alpha \left. \{ x^\mu , x^\nu \} \right \vert _{ x' (x) }$ for some $\alpha \in \C$.
For such theories, let $x^\mu (t)$ be a dynamical trajectory for theory 1, and define $\widetilde x ^\mu (t) = x' {}^\mu \left( x(t) \right)$.
Then one can check that $\dot{ \widetilde x } {}^\mu =\lambda_2 \{  \widetilde x^\mu , H_2 \}$ for $\lambda_2 \left( x' (x) \right) = \alpha \lambda_1 (x)$, so that $\widetilde x ^\mu (t)$ is a dynamical trajectory for theory 2 with this particular choice of $\lambda_2$:
\begin{align}
        \dot{ \widetilde x } {}^\mu& = \left. \left( \frac{ \partial x' {}^\mu }{ \partial x^\nu } \right) \right \vert_{ x(t) } \dot x^\nu (t)
        = \left. \left( \lambda_1 \{ x^\nu , H_1 \} \, \frac{ \partial x' {}^\mu }{ \partial x^\nu } \right) \right \vert_{ x(t) }
        = \lambda_1 \left( x(t) \right) \left. \{ x' {}^\mu , x' {}^\nu \} \right \vert_{ x(t) } \left. \left( \frac{ \partial H_2 }{ \partial x^\nu } \right) \right \vert_{ \widetilde x (t) }\nonumber
        \\
        & = \alpha \lambda_1 \left( x(t) \right) \left. \left( \{ x^\mu , x^\nu \} \, \frac{ \partial H_2 }{ \partial x^\nu } \right) \right \vert_{ \widetilde x (t) }
        = \lambda_2 \left( \widetilde x (t) \right) \{  \widetilde x^\mu , H_2 \} \,.
\end{align}

The Euclidean theory can be connected to all the other Lorentzian solution branches by similar ``Wick rotations''. For instance, the Lorentzian Hamiltonian constraint (\ref{Real Hamiltonian constraint 2real 1imaginary}), corresponding to a case in which the surfaces of homogeneity are timelike, is obtained from the Euclidean Hamiltonian constraint $\mathcal{H}_\text{E}$ after a transformation $\gamma^3\mapsto \ii\gamma^3\,,\;p_1\mapsto\ii p_1\,,\;p_2\mapsto\ii p_2$ with all other variables unchanged. Obviously such a transformation has $\{\cdot,\cdot\}\mapsto \ii\{\cdot,\cdot\}$ and so again fits into the scope of our general result. 

One may then take the view, in particular when looking at quantisation, that the fundamental definition of the chiral connection formulation of general relativity should be the Euclidean theory. Lorentzian solutions emerge from such a theory for purely imaginary boundary conditions, with the different branches we found earlier corresponding to different types of variables chosen as real or imaginary. In the case of cosmological solutions, where the transformations we have discussed are defined unambiguously, this starting point might be preferable over dealing with the full complex theory and its reality conditions. As in the conventional metric formulation, it seems much less clear in which sense such a correspondence extends to the full theory.

\section{Homogeneous and isotropic case}
\label{homoiso}
 
In section \ref{Diagonal symmetric models section} we saw that the Hamiltonian dynamics of the diagonal Bianchi IX model, for solutions corresponding to real Lorentzian signature metrics for which the homogeneous hypersurfaces are spacelike, can be defined in terms of a Hamiltonian constraint on a phase space $\mathcal{P}_\text{R}=\R^6$ with Poisson bracket $\{ \gamma^i , p_j \} = \delta^i _j$. The Hamiltonian constraint can be written as $\mathcal{H} = \mathcal{H}_{ ( - ) } + \kappa^2 \, \mathcal{H}_\text{BG}$, see (\ref{Real Hamiltonian constraint}), and the Hamiltonian is $\lambda\mathcal{H}$ where $\lambda$ is a Lagrange multiplier specifying the time coordinate.

The expressions for the self-dual connection and Urbantke metric in terms of these dynamical variables are (cf.~(\ref{Urbantke metric}), (\ref{field definitions Bianchi IX}))
\begin{align}
 & A^i = V^{-1/3} \left( \ii \gamma^i + \Pi ^i (p) \right) e^i\,, \qquad \Pi^1 (p) := \frac{ \kappa p_2 p_3 }{ 2 } \left( - \frac{ 1 }{ (p_1)^2 } + \frac{ 1 }{ (p_2)^2 } + \frac{ 1 }{ (p_3)^2 } \right)\quad \text{etc.}\,,
\\& g_\Sigma = - \frac{ \ell_\text{P}^2\, \lambda^2 }{ p_1 p_2 p_3 } \, \dee t \otimes \dee t + V ^{-2/3} \ell_\text{P}^2 \, p_1 p_2 p_3 \, \sum_{i=1}^3 \frac{1}{ (p_i)^2 } \, e^i \otimes e^i\,.
\end{align}

We now want to discuss the specific case where the metric is also isotropic, i.e., of FLRW form. We can see that this case corresponds to $p_1=p_2=p_3$ for all $t$. One can check that this restriction evolves consistently only if we also have $\gamma^1=\gamma^2=\gamma^3$. Hence, starting from the fields appearing in the Bianchi IX model, we can make the substitutions $\gamma^i\rightarrow c/3$ and $p_i\rightarrow p$ where $c(t)$ and $p(t)$ are real scalars. This yields a connection and curvature
\be
A^i = V^{-1/3} \left( \frac{\ii c}{3} + \frac{\kappa}{2} \right) e^i\,,\qquad F^1 = \ii V^{-1/3}\frac{\dot{c}}{3}\,\dee t\wedge e^1 - V^{-2/3}\frac{c^2+K}{9}e^2\wedge e^3\quad\text{etc.}
\ee
where $K:=(3\kappa/2)^2=9V^{2/3}k>0$ is a rescaled spatial curvature parameter. The curvature 4-form $F^i\wedge F^j$ is now proportional to $\delta^{ij}$, so that we can also take $M_{ij}\propto \delta_{ij}$. Indeed, the expression for the auxiliary matrix in (\ref{field definitions Bianchi IX}) is now
\be
M^{ij} = \frac{c^2+K}{9\,\ell_\text{P}^2\,p}\,\delta^{ij}\,.
\ee
We can now obtain an action for the FLRW Universe either by substituting these ans\"atze into the original action (\ref{Chiral first order action}) or from the Bianchi IX reduced action (\ref{Reduced Lorentzian action}). In either case, the result is
\begin{equation}
\label{FLRW action}
    S_\text{Iso} \left[ c , p , \bar\lambda \right] =  \int \dee t\left(p \, \dot c - \lambda \left(   \Lambda _\text{P} - \frac{ c^2+K}{ 3p }\right)\right)=
 \int \dee t\left(p \, \dot c - \bar\lambda \left( p-\frac{c^2+K}{3\Lambda_\text{P}}\right)\right)
\end{equation}
where we are redefining the Lagrange multiplier as $\bar\lambda=\frac{\Lambda_\text{P}}{p}\lambda$.
The Euler--Lagrange equations resulting from this action are
\be
\label{FLRW equations}
\dot{c}=\bar\lambda\,,\qquad \dot{p}=2\bar\lambda\,\frac{c}{3\Lambda_\text{P}}\,.
\ee
These are the same equations that one would obtain from (\ref{eqmo1}) and (\ref{eqmo2}) after the substitutions $C^i=\frac{c}{3}-\ii\frac{\kappa}{2}$, $P_i=p$ and $\lambda=\frac{p}{\Lambda_\text{P}}\bar\lambda$. Hence the restriction to isotropy commutes with the variational principle. The Urbantke metric now takes the explicit form
\be
g_\Sigma = - \frac{ \ell_\text{P}^2\, \bar\lambda^2 }{\Lambda_\text{P}^2\, p } \, \dee t \otimes \dee t + V ^{-2/3} \ell_\text{P}^2 \, p \, \sum_{i=1}^3  e^i \otimes e^i\,.
\ee
This parametrisation of the $k>0$ FLRW metric is well-known in quantum cosmology, see, e.g., the expression given in \cite{Halliwell},
\be
\dee s^2=-\frac{N^2(t)}{q(t)}\,\dee t^2+q(t)\,\dee\Omega_3^2
\ee
and subsequent discussion of the path integral quantisation given there. (Note the variable $q$ used in \cite{Halliwell} corresponds to our $p$, whereas the variable $p$ used there is conjugate to $q$, i.e., corresponds to the connection $c$ in our notation.)

The definition of a path integral for this system depends in general on the choice of boundary conditions as discussed, e.g., in \cite{LehnersBound}. The path integral in \cite{Halliwell} corresponds to the case where one wants to keep the spatial metric fixed at the initial and final times, so that in our variables $p$ would be fixed. One is then interested in an amplitude or two-point function of the form
\be
G(p_f|p_i)=\int \mathcal{D}c\;\mathcal{D}p\;\mathcal{D}\bar\lambda\;\exp\left[\ii \int_{t_i}^{t_f} \dee t\left(p \, \dot c - \bar\lambda \left( p-\frac{c^2+K}{3\Lambda_\text{P}}\right)\right)\right]
\ee
where the allowed paths for $p$ must start at $p_i$ and end at $p_f$. Due to the gauge symmetry of the system under reparametrisations in time, this integral is ill-defined as given. The standard approach proposed in \cite{Halliwell} (and also followed in \cite{LorentzianQC}) is to impose a gauge fixing, such as $\dot{\bar\lambda}=0$, by adding a term to the action. To ensure that the final result does not depend on the gauge choice, one can then introduce additional anticommuting ghost fields to make the action invariant under a global Becchi--Rouet--Stora--Tyutin (BRST) symmetry. The path integral over the ghost fields can be done explicitly and leads to a new expression for the gauge-fixed path integral,
\be
G(p_f|p_i)=\int \dd\bar\lambda(t_f-t_i)\int \mathcal{D}c\;\mathcal{D}p\;\exp\left[\ii \int_{t_i}^{t_f} \dee t\left(p \, \dot c - \bar\lambda \left( p-\frac{c^2+K}{3\Lambda_\text{P}}\right)\right)\right]\,.
\ee
There is no longer a path integral over the Lagrange multiplier $\bar\lambda$, just an ordinary integral whose definition depends on what kind of two-point function one is interested in \cite{HalliwellOrtiz}. In the case where one is interested in solutions to the canonical Wheeler--DeWitt equation that may be interpreted as physical wavefunctions, one possible integration contour for $\bar\lambda$ is over the entire real line, leading to the usual ``no-boundary'' solutions expressed in terms of Airy functions \cite{Halliwell}.

In the context of a pure connection approach to gravity, it seems more natural to specify the connection $c$ at the initial and final times and hence try to compute the path integral
\be
G(c_f|c_i)=\int \dd\bar\lambda(t_f-t_i)\int \mathcal{D}c\;\mathcal{D}p\;\exp\left[\ii \int_{t_i}^{t_f} \dee t\left(p \, \dot c - \bar\lambda \left( p-\frac{c^2+K}{3\Lambda_\text{P}}\right)\right)\right]
\ee
using the same gauge-fixing method as before. This is now rather straightforward, since the $p$ path integral may be defined through a time-slicing as
\be
 \int\mathcal{D}p\;\exp\left[\ii \int_{t_i}^{t_f} \dee t\left(p \, \dot c - \bar\lambda p\right)\right]
= \lim_{N\rightarrow\infty}\prod_{i=1}^N \int \frac{\dee p_i}{2\pi}\;e^{\ii p_i(c_i-c_{i-1}-\bar\lambda\,\delta t)}=\lim_{N\rightarrow\infty}\prod_{i=1}^N \delta(c_i-c_{i-1}-\bar\lambda\,\delta t)
\ee 
with $c_0=c_i$ and $c_N=c_f$ and $\delta t=\frac{t_f-t_i}{N}$. We are then left with
\begin{align}
G(c_f|c_i)&=\lim_{N\rightarrow\infty}\int \dd\bar\lambda(t_f-t_i)\int \dee c_1\ldots\dee c_{N-1}\;e^{\ii\bar\lambda\sum_{l=1}^N\delta t\frac{c_l^2+K}{3\Lambda_\text{P}}} \prod_{i=1}^N \delta(c_i-c_{i-1}-\bar\lambda\,\delta t)\nonumber
\\&=\lim_{N\rightarrow\infty}\int \dd\bar\lambda(t_f-t_i)\;\delta(c_f-c_i-\bar\lambda(t_f-t_i))\,e^{\ii\bar\lambda\sum_{l=1}^N\delta t\frac{(c_i+l\bar\lambda\delta t)^2+K}{3\Lambda_\text{P}}} \nonumber
\\&=\lim_{N\rightarrow\infty}\exp\left[\frac{\ii}{N}(c_f-c_i)\sum_{l=1}^N\frac{(c_i+\frac{l}{N}(c_f-c_i))^2+K}{3\Lambda_\text{P}}\right]\nonumber
\\&=\exp\left[\ii\frac{\frac{1}{3}(c_f^3-c_i^3)+K(c_f-c_i)}{3\Lambda_\text{P}}\right]
\label{path integral calc}
\end{align}
where the remaining delta function in the second line requires $\bar\lambda=\frac{c_f-c_i}{t_f-t_i}$ and we assume that the chosen contour for $\bar\lambda$ is the real line. The path integral with $c$ boundary conditions can hence be done analytically and yields a simple result of pure plane wave form, which may be written as
\be
G(c_f|c_i)=\psi_{{\rm CS}}(c_f)\bar{\psi_{{\rm CS}}}(c_i)
\ee
where $\psi_{{\rm CS}}(c):=\exp(\frac{\ii}{3\Lambda_\text{P}} \left(\frac{1}{3}c^3+K\,c\right))$ is the unique (up to normalisation) solution to the Wheeler--DeWitt equation
\be
 \ii \frac{\dee}{\dee c}\psi(c)=-\frac{c^2+K}{3\Lambda_\text{P}}\psi(c)
\ee
corresponding to the classical Hamiltonian constraint appearing in (\ref{FLRW action}). $\psi_{{\rm CS}}(c)$ can be seen as the restriction of a ``Chern--Simons state'' that can be defined in more general situations \cite{ChernSimons} to homogeneous and isotropic Universes  (see also \cite{RealCS} for generalisations of the Chern--Simons state). The state can be seen as related to the Hartle--Hawking or Vilenkin wavefunctions of the metric formulation via a kind of Fourier transform \cite{MagueijoDuality}.

We see that the two-point function $G(c_f|c_i)$ with connection boundary data is straightforward to obtain, as also shown recently in \cite{RayJoao} following a very similar calculation. The result has an interesting, and somewhat novel, interpretation from the perspective of pure connection formulations of general relativity investigated in this paper. Notice that starting from the classical action (\ref{FLRW action}), one may ``integrate out'' the field $p$ by substituting the solution to the Hamiltonian constraint,
\be
\label{FLRW constraint solution}
 p=\frac{c^2+K}{3\Lambda_\text{P}}\,,
\ee
back into the Lagrangian. This yields
\be
\label{pure boundary action}
    S_\text{Iso} =
 \int \dee t\left(\frac{c^2+K}{3\Lambda_\text{P}}\dot c\right) = \left[\frac{\frac{1}{3}c^3+K\,c}{3\Lambda_\text{P}}\right]^{t_f}_{t_i}\,,
\ee
i.e., the Lagrangian is a total derivative and the action becomes a pure boundary term.\footnote{This property follows directly from the fact that here the Weyl curvature vanishes, and so the auxiliary matrix $M_{ij}$ is (on-shell) proportional to $\delta_{ij}$. We thank Kirill Krasnov for pointing this out.} There are hence no non-trivial equations of motion left for $c$, but this is as expected, given that (\ref{FLRW equations}) shows that {\em any} function $c(t)$ is a solution in a certain gauge, given by $\dot{c}=\bar\lambda$.

The reduced action (\ref{pure boundary action}) is of course nothing but the reduction of the chiral pure connection action (\ref{Pure connection action}) to a homogeneous and isotropic connection. Indeed, given that
\be
F^i\wedge F^j = 2\ii\,\delta^{ij} \,\frac{\dot{c}(c^2+K)}{27V}\,e^1\wedge e^2\wedge e^3\wedge \dee t\,,
\ee
we may choose the top-form $\varepsilon_X= e^1\wedge e^2\wedge e^3\wedge \dee t$ and define the matrix $X^{ij}$ appearing in (\ref{Pure connection action}) as
\be
X^{ij} = \delta^{ij} \,\frac{\dot{c}(c^2+K)}{27V}\,.
\ee
We then find a pure connection action
\begin{equation}
    S_{{\rm PC}} [ c ] = \frac{1}{ \Lambda_\text{P} } \int \left( \tr \sqrt{ X } \right)^2 \varepsilon_X = \frac{V}{\Lambda_\text{P}}\int \dee t\,\frac{\dot{c}(c^2+K)}{3V}
\end{equation}
in agreement with (\ref{pure boundary action}): for FLRW geometries the chiral pure connection action is a pure boundary term, consistent with the fact that this system has no dynamical degrees of freedom and, given that there is no Lagrange multiplier related to time reparametrisations in this theory, there cannot be any nontrivial dynamical equations.

An attempt to define a path integral directly at the Lagrangian level for the pure connection formulation would lead to an expression
\be
G(c_f|c_i) = \int\mathcal{D}c\;\exp\left(\ii S_{{\rm PC}}[c]\right)
\ee
but since $S_{{\rm PC}}[c]$ is a pure boundary term, one would again be left with a divergent integration over redundant, gauge-equivalent configurations, which needs gauge fixing to be well-defined.  Any such gauge fixing will turn the integration over $c$ into a constant factor leading to our previous result (\ref{path integral calc}). The case of homogeneous isotropic connections is hence one in which the pure connection path integral has an immediate exact definition given in terms of the classical action,
\be
G(c_f|c_i)=\exp\left(\ii S_{{\rm PC}}(c_f,c_i)\right)\,.
\ee

One may now discuss particular choices of $c_f$ and $c_i$, in particular ``no-boundary'' conditions for $c_i$. Recall that the general idea as pioneered by Hartle and Hawking is that the initial state of the Universe would be described by a Euclidean 4-sphere appearing to emerge from zero size (but such that the resulting geometry is actually regular). Such an initial condition is often interpreted as corresponding to zero scale factor, or $p=0$ in our notation. However, it has been suggested (e.g., in \cite{LehnersBound}, building on earlier work such as \cite{LoukoCanon}) that an initial condition should rather be put on the connection to distinguish between different semiclassical saddle point solutions, and in particular single out the ``Hartle--Hawking'' over the ``Vilenkin'' solution. Given the constraint (\ref{FLRW constraint solution}), a no-boundary initial condition on the connection corresponds to $c=\pm\ii\sqrt{K}$. Which such an initial condition the two-point function of interest becomes
\be
G(c_f|c_i)=\psi_{{\rm CS}}(c_f)\bar{\psi_{{\rm CS}}}(\pm\ii\sqrt{K})=\exp\left(\mp\frac{2}{9}\frac{K^{3/2}}{\Lambda_\text{P}}\right)\psi_{{\rm CS}}(c_f)=\exp\left(\mp \frac{6Vk^{3/2}}{\Lambda_\text{P}}\right)\psi_{{\rm CS}}(c_f)
\ee
so that depending on the choice of sign we get either exponential suppression ({\em \`a la} Vilenkin) or exponential enhancement ({\em \`a la} Hartle--Hawking). Inserting the usual choices $k=1$ and $V=2\pi^2$ this factor is $\exp\left(\mp \frac{12\pi^2}{\Lambda_\text{P}}\right)$, consistent with the literature \cite{QCdebate}.

\section{Conclusions}

Our study of homogeneous cosmological models in pure connection formulations of general relativity with cosmological constant $\Lambda$ has exemplified various conceptual and technical differences of these formulations when compared with approaches based on a metric or (Ashtekar--Barbero) connection and tetrad. 

Restriction to spatial homogeneity means that the variables of the gravitational field are reduced from free functions of space and time to only finitely many free functions of time. In particular, in diagonal Bianchi IX and Bianchi I models the connection has only three independent components. We could isolate the canonically conjugate variables in the ``auxiliary matrix''; in the first order chiral connection formulation that we have mostly focused on, these variables can be interpreted as defining a metric through the Urbantke construction, such that this reconstructed metric satisfies the Einstein equations. A crucial issue in these formulations is that all fields are initially complex-valued, and Lorentzian solutions are only obtained from imposing reality conditions. The solutions to these reality conditions fall into different classes, some with a Lorentzian Urbantke metric (either of signature $(-+++)$ or $(+---)$) and some with an imaginary Urbantke metric. The latter cases would require extra care in defining a Lorentzian metric, but for real (and non-zero) $\Lambda$ they do not have any consistent solutions to the field equations, which is why we could mostly ignore them. (The fact that these cases do not admit consistent solutions is independent of the definition of the Urbantke metric, which is not a fundamental variable of the theory.)

In the general complex case, the reduced dynamics of cosmological models take the form of a classical particle system in complex space. The dynamics can be expressed in the form of a holomorphic Hamiltonian theory over a complex phase space whose Hamiltonian is constrained to vanish. However, the reality conditions needed to obtain Lorentzian solutions are neither holonomic nor holomorphic. To deal with the former issue, the dynamical equations could be used to eliminate time derivatives leaving us with algebraic conditions on the variables. To deal with the latter issue, we constructed a new Hamiltonian system over real variables, starting from either the real or imaginary part of the full, generally complex, action. The reality conditions then split into four possible cases, each of which needs to be dealt with separately. These conditions and their associated secondary consistency conditions can be viewed as second class constraints in the Dirac programme; we derived the Dirac bracket and associated phase space. Viewing the second class constraints as strong conditions allowed us to eliminate half of the dynamical variables. Different branches of solutions to the reality conditions can reproduce the usual Bianchi models, in which the homogeneous surfaces are spacelike, or models in which these surfaces are timelike and the direction of ``evolution'' is spacelike. Focusing on the physically more relevant first case, we explicitly recovered the Lorentzian Bianchi IX model. We then repeated the analysis for the Bianchi I model, with similar results and analytical solutions reproducing the known Kasner solutions.

The Pleba\'{n}ski formulation of gravity and its descendants allow a definition in Euclidean signature in terms of real variables only, so that no reality conditions are required. This situation is rather different from other formulations of gravity in which the Euclidean and Lorentzian theories have the same number of dynamical variables and constraints. We showed that the Euclidean theory of the Bianchi IX model allows complex transformations into all possible Lorentzian branches studied earlier. These transformations may be interpreted as a type of Wick rotation, suggesting that the Euclidean theory may be a more natural starting point for defining the theory especially in the quantum regime, where reality conditions will be awkward to handle. The signature ambiguity however persists: the Euclidean Urbantke metric can have signature $(++++)$ or $(----)$.

In the last section, we examined the isotropic limit of the Bianchi IX model which corresponds to the FLRW spacetime with closed spatial surfaces.  This restriction leaves only a single dynamical pair of variables subject to a constraint, with the remaining degree of freedom pure gauge. We showed that the two-point function between states with connection boundary data, defined by a suitable path integral, reduces to $e^{\ii S}$ where $S$ is a pure boundary term. This result has been derived before \cite{RayJoao} but we gave it a somewhat novel classical interpretation: we saw that in the FLRW model one can explicitly integrate out all variables apart from the connection, and obtain a classical action which is a pure boundary term. This action is the reduction of the pure connection formulation of \cite{KrasnovPureConn} to the FLRW model which hence has an immediate, trivial, quantum definition as a path integral. One of the main questions for future work would be how much these drastic simplifications achieved in the pure connection formulation can be extended to more complicated models such as Bianchi models or perhaps black hole spacetimes.

As we have discussed, an interesting nuance of the chiral connection formulations of gravity is the inability to fix the signature of the metric throughout.  Even in a specific branch of solutions to the reality conditions, one obtains two types of solutions with different metric signature, unlike in metric-based formulations or the Ashtekar--Barbero approach where the metric signature is fixed. Dynamical signature change within a solution would require passing through a surface with degenerate metric and divergent curvature, which we may see as a classical endpoint of a given solution. But in any case, this implies that $\Lambda$ in our original action (\ref{Chiral first order action}) represents the cosmological constant of general relativity only {\em up to sign}: the Urbantke metric in the chiral connection framework satisfies the Einstein equations $R_{\mu\nu}=\Lambda g_{\mu\nu}$, but its signature cannot be fixed, and a change of signature $g_{\mu\nu}\rightarrow -g_{\mu\nu}$ could be absorbed in $\Lambda\rightarrow -\Lambda$. In the homogeneous and isotropic sector, we always find both de Sitter and anti-de Sitter solutions. This fact has interesting implications if we wanted to fix the value of $\Lambda$ in (\ref{Chiral first order action}) by comparing with observation; even observing accelerated expansion would only determine the magnitude of $\Lambda$, not its sign. These conclusions are somewhat reminiscent of arguments in favour of the emergence of expanding solutions for negative $\Lambda$ in quantum cosmology \cite{HartleHawkingHertog}, but here appear already in the classical theory. They seem to be a feature only of chiral connection formulations of gravity, which is {\em essential} for the existence of a ``pure connection'' formulation in which all other variables have been integrated out.  Indeed, in approaches with fixed metric signature (\ref{FLRW constraint solution}) would usually be a condition on $|p|$ rather than $p$ (or $p$ would be fixed to be positive, unlike what we observe here), so that we would not obtain (\ref{pure boundary action}). We hope to explore these fascinating questions more in future work.
\\
\\{\bf Acknowledgments:} We thank Jo\~ao Magueijo for discussions related to this work, and Kirill Krasnov for comments on the manuscript. The work of SG was funded by the Royal Society through a University Research Fellowship (UF160622).

\appendix
\section{Holomorphic Lagrangian mechanics}
\label{Appendix complex lagrangian}
Consider an action of the form
\begin{equation}
    S [ Q ] = \int \dee t \; L \left( Q , \dot Q \right)
\end{equation}
where $Q^i (t)$ are complex variables and the Lagrangian $L(Q , \dot Q)$ is also complex valued.
We assume that $L$ is holomorphic so that
\begin{equation}
    \label{holomorphic condition}
    \frac{ \partial L }{ \partial \overline{Q} {}^i } = 0\,,\quad
    \frac{ \partial L }{ \partial \dot{ \overline{Q} } {}^i } = 0\,.
\end{equation}
We can decompose the complex variables into their real and imaginary parts as $Q^i = x^i + \ii y^i$. Then the Lagrangian can be decomposed as $L (Q , \dot Q) = L_R \left( x , y , \dot x , \dot y \right) + \ii L_I \left( x , y , \dot x , \dot y \right)$ where $L_R$ and $L_I$ are real valued functions of real variables.
Using the complex partial derivatives (Wirtinger derivatives), the holomorphic conditions (\ref{holomorphic condition}) become the Cauchy--Riemann equations
\begin{equation}
    \frac{ \partial L_R }{ \partial x^i } = \frac{ \partial L_I }{ \partial y^i }\,, \quad
    \frac{ \partial L_I }{ \partial x^i } = - \frac{ \partial L_R }{ \partial y^i }\,,\quad
    \frac{ \partial L_R }{ \partial \dot x ^i } = \frac{ \partial L_I }{ \partial \dot y ^i }\,,\quad
    \frac{ \partial L_I }{ \partial \dot x ^i } = - \frac{ \partial L_R }{ \partial \dot y ^i }\,.
\end{equation}
For a functional $F$ over real variables $z^a (t)$ defined $F [z] = \int \dee t \; f ( z , \dot z )$ for some function $f$, the functional derivatives are 
\begin{equation}
    \frac{ \delta F }{ \delta z^a } = \frac{ \partial f }{ \partial z^a } - \frac{\dee}{\dee t} \left( \frac{ \partial f }{ \partial \dot z ^a } \right)\,.
\end{equation}
Using this expansion one obtains functional Cauchy--Riemann equations
\begin{equation}
    \label{functional euler lagrange eqns}
    \frac{ \delta L_R }{ \delta x^i } = \frac{ \delta L_I }{ \delta y^i }\,,\quad
    \frac{ \delta L_I }{ \delta x^i } = - \frac{ \delta L_R }{ \delta y^i }\,.
\end{equation}
In the notation used in the main text, the real part of the action $S_R [ x . y ]$ generates the ``real branch'' of the theory while the imaginary part $S_I [ x , y ]$ generates the ``imaginary branch''.
The Euler--Lagrange equations of the real and imaginary branches are
\begin{subequations}
    \begin{align}
        \label{real branch el eqns}
        \text{Real Branch}
        \quad & : \quad
        \frac{ \delta L_R }{ \delta x^i } = 0\,, \quad \frac{ \delta L_R }{ \delta y^i } = 0\,;
        \\[0.5em]
        \label{imaginary branch el eqns}
        \text{Imaginary Branch}
        \quad & : \quad
        \frac{ \delta L_I }{ \delta x^i } = 0 \,, \quad \frac{ \delta L_I }{ \delta y^i } = 0\,.
    \end{align}
\end{subequations}
From (\ref{functional euler lagrange eqns}) we then see that the Euler--Lagrange equations (\ref{real branch el eqns}) and (\ref{imaginary branch el eqns}) are equivalent, and one may focus on only one of them. The discussion here is similar to that in Section 5 of \cite{KrasnovMitsou}.

\section{Hamiltonian constraint for the Lorentzian Bianchi IX model}
\label{bianchiIXapp}

Here we show explicitly that (\ref{Real Hamiltonian constraint}) reproduces the dynamics of the Lorentzian Bianchi IX model in general relativity, as discussed in \cite{WilsonEwingBianchiIX} in the context of Ashtekar--Barbero formulation. This formulation is based on an $SU(2)$ connection defined on each spatial hypersurface in a given foliation as $A^i=\Gamma^i[E]+\beta K^i$ where $\Gamma^i$ is the torsion-free (Levi-Civita) connection 1-form associated to a given triad $E$, $K^i$ is the extrinsic curvature 1-form, and $\beta$ is a free parameter generally taken to be real. The Bianchi IX ansatz made in \cite{WilsonEwingBianchiIX} is
\be
A^i=V^{-1/3} c^i\,\tilde{e}^i\,,\quad E_i=V^{-2/3}\pi_i\,\sqrt{\tilde{q}}\,{\bf e}_i
\ee
where $E_i$ is the ``densitised triad'' conjugate to $A^i$, ${\bf e}_{\bf a}$ is a co-frame of vector fields dual to the frame $\tilde{e}^{\bf a}$, and $\sqrt{\tilde{q}}$ is the volume element associated to the ``fiducial metric'' $\tilde{q}:=\delta_{{\bf a}{\bf b}}\tilde{e}^{\bf a}\otimes \tilde{e}^{\bf b}$. The frame $\tilde{e}^{\bf a}$ is assumed to satisfy $\dd \tilde{e}^{\bf a}-\frac{1}{r_o}\, \epsilon^{\bf a} {}_{{\bf b}{\bf c}} \, \tilde{e}^{\bf b} \wedge \tilde{e}^{\bf c} = 0$. The variables $c^i$ and $\pi_i$ then satisfy
\be
\{c^i,\pi_j\}=\beta\,\ell_\text{P}^2\,\delta^i_j\,.
\ee
Recall that on the branch of the reality conditions that leads to (\ref{Real Hamiltonian constraint}), the self-dual connection of the chiral connection theory can be written as $A^i = V ^{-1/3} (\ii\gamma^i-\Gamma^i) \, e^i$ with $\Gamma^i$ fixed by (\ref{Lagrangian secondary constraints}) also corresponding to the Levi-Civita connection on spatial hypersurfaces. We can relate this complex connection to the Ashtekar--Barbero connection by setting $\beta=-\ii$, the choice made in the original Ashtekar formulation \cite{Ashtekar}. Our Cartan frame $e^{\bf a}$ is also defined with the opposite orientation, so in order to relate to \cite{WilsonEwingBianchiIX} we need to identify $e^{\bf a}=-\tilde{e}^{\bf a}$ with $\sqrt{k}=1/r_o$.

With $\beta=-\ii$, the expression for the (pure gravity) Hamiltonian given in \cite{WilsonEwingBianchiIX} becomes
\be
N\mathcal{C}_H = \frac{N}{\ell_\text{P}^2\sqrt{|\pi_1\pi_2\pi_3|}}\left(\pi_1\pi_2c_1c_2+\pi_1\pi_3c_1c_3+\pi_2\pi_3c_2c_3+\kappa\left(\pi_1\pi_2c_3+\pi_2\pi_3c_1+\pi_1\pi_3c_2\right)\right)
\ee
where the orientation factor is given by $\varepsilon=+1$. We now need to rewrite this Hamiltonian in terms of the variables used in the main text. First of all, we can identify
\be
c^i = -\ii\gamma^i-\Pi^i(p)\,,\quad \pi_i = \ell_\text{P}^2\, p_i
\ee
where the first equation follows from our identification of the real and imaginary parts of the self-dual connection, and the second is a rescaling to ensure canonical Poisson brackets $\{\gamma^i,p_j\}=\delta^i_j$ (which may also be obtained from equating the Urbantke metric with the physical metric defined in \cite{WilsonEwingBianchiIX}). Finally, using the Urbantke metric (\ref{Urbantke metric}), we can write the lapse $N$ as $N = \ell_\text{P} |p_1 p_2 p_3|^{-1/2} \lambda$ in terms of the Lagrange multiplier $\lambda$. We then find
\be
\label{GRHamiltonian}
N\mathcal{C}_H = \lambda\,{\rm sgn}(p_1 p_2 p_3)\left(-\frac{\gamma^1 \gamma^2}{p_3}-\frac{\gamma^1 \gamma^3}{p_2}-\frac{\gamma^2 \gamma^3}{p_1}+\frac{\kappa^2}{2}\left(\frac{p_1 p_2}{2p_3^3}+\frac{p_1 p_3}{2p_2^3}+\frac{p_2 p_3}{2p_1^3}-\frac{p_1}{p_2 p_3}-\frac{p_2}{p_1 p_3}-\frac{p_3}{p_1 p_2}\right)\right)
\ee
which equals (apart from the overall sign) the purely gravitational part of the Hamiltonian $\lambda\,\mathcal{H}=\lambda\left(\mathcal{H}_{ ( - ) } + \kappa^2 \, \mathcal{H}_\text{BG}\right)$ in (\ref{Real Hamiltonian constraint}). The contribution of the cosmological constant replaces the matter Hamiltonian for a massless scalar field appearing in \cite{WilsonEwingBianchiIX}, specifically
\be
\mathcal{H}_{matt} = \frac{N}{2}\,\frac{p_T^2}{\sqrt{|q|}}\quad\rightarrow\quad\frac{N}{\ell_\text{P}^2}\Lambda\,\sqrt{|q|}
\ee
which means adding an extra term
\be
\frac{N}{\ell_\text{P}^2}\Lambda\,\sqrt{|\pi_1\pi_2\pi_3|} = \lambda\,\ell_\text{P}^2\Lambda
\ee
to (\ref{GRHamiltonian}).  This is exactly the $\Lambda$ term found in (\ref{Real Hamiltonian constraint}), which shows that the Hamiltonian dynamics recover those of the Lorentzian Bianchi IX model with a given $\Lambda$ for $p_1 p_2 p_3 > 0$. For the opposite sign, from (\ref{GRHamiltonian}) one would expect to see a relative minus sign which is not seen in (\ref{Real Hamiltonian constraint}). This discrepancy is explained by the fact that in that case the Urbantke metric (\ref{Urbantke metric}) has signature $(+---)$, but satisfies the Einstein equations with the same $\Lambda$; it is hence equivalent to a solution with signature $(-+++)$ but cosmological constant $-\Lambda$. The Ashtekar formulation of general relativity assumes signature $(-+++)$ throughout and can match the sign only for $p_1 p_2 p_3 > 0$.

\end{document}